\documentclass{aa}

\usepackage{graphics}
\usepackage{epsfig}
\def\bib{\bibitem{}}

\newcommand{\xia}{\overline{\xi}}
\newcommand{\rhoa}{\overline{\rho}}
\newcommand{\rhob}{\overline{\rho}}

\newcommand{\gam}{\gamma}
\newcommand{\pl}{\partial}

\newcommand{\beq}{\begin{equation}}
\newcommand{\eeq}{\end{equation}}
\newcommand{\lag}{\langle}
\newcommand{\rag}{\rangle}
\begin{document}
%
%
%
%

\thesaurus{Sect.02 (12.12.1; 11.05.2; 11.17.3; 11.09.3; 11.03.1)} 
\title{The entropy history of the universe.}
\author{Patrick Valageas\inst{1} \and Joseph Silk\inst{2,}\inst{3,}\inst{4}}
\institute{Service de Physique Th\'eorique, CEA Saclay, 91191 
Gif-sur-Yvette, France 
\and
Center for Particle Astrophysics, Department of Astronomy and Physics,
University of California, Berkeley, CA 94720-7304, USA
\and
Astrophysics, Department of Physics, Keble Road, Oxford OX1 3RH, U.K.
\and
Institut d'Astrophysique de Paris, CNRS, 98bis Boulevard Arago,
F-75014 Paris, France}
\date{Received / Accepted }
\maketitle 
\markboth{P. Valageas \& J. Silk: The entropy history of the universe}{P. Valageas \& J. Silk: The entropy history of the universe}

\begin{abstract}

Although usual hierarchical clustering scenarios agree with the main observed 
properties of Lyman-$\alpha$ clouds and galaxies, they seem to fail to 
reproduce the cluster temperature - X-ray luminosity relation. A possible 
explanation is that {\it the IGM is reheated by supernovae or quasars before cluster 
formation}. In this article, using a unified analytic model for quasars, 
galaxies, Lyman-$\alpha$ absorbers and the IGM, we obtain the redshift 
evolution of the temperature and the entropy of the gas and the corresponding cluster temperature - X-ray luminosity relation. We consider three 
scenarios: 1) no energy source in addition to photoionization heating, 2) 
heating from supernovae and 3) heating from quasars, in an open universe 
$\Omega_m=0.3$, $\Omega_{\Lambda}=0$, with a CDM power-spectrum. We show that although 
{\it quasars can easily reheat the IGM} and raise its entropy up to the level 
required by current cluster observations {\it the energy provided by supernovae is 
unlikely to be sufficient}. Indeed, the efficiency factor needed for the 
supernova scenario is of order unity ($\alpha_{SN} \simeq 
1.7$) while for quasars we get $\alpha_Q \simeq 0.008$. Thus the IGM is more 
likely to have been reheated by quasars. Moreover, we find that if both 
scenarios are normalized to present observations the reheating due to quasars 
occurs somewhat earlier ($z_S \sim 2$) than for supernovae ($z_S \sim 0.4$) 
because of the sharp drop at low $z$ of the quasar luminosity function. We also 
show that the Compton parameter $y$ induced by the IGM is well below the 
observed upper limit in all cases. Finally, we note that {\it such a reheating 
process may partly account for the decline at low redshift of the comoving 
star formation rate} and of the quasar luminosity function. In particular, we 
show that the contradictory requirements arising from clusters (which require 
a large reheating to affect the relation $T-L_X$) and galaxies (which require 
a small reheating so that galaxy and star formation are not too much inhibited) 
provide strong constraints on such models. Thus, the IGM should be reheated at 
low redshift $z \la 2$ up to $T \sim 5 \; 10^5$ K. On the other hand, the 
reionization process of the universe is almost not modified by these entropy 
sources which means that our predictions for the former should be quite robust.

\end{abstract}

\keywords{cosmology: large-scale structure of Universe - galaxies: 
evolution - quasars: general - intergalactic medium - galaxies: clusters}

\section{Introduction}

An important test of cosmological models is to check whether they can reproduce the wide variety of astrophysical objects we observe in the present universe. Several studies have already shown that the usual hierarchical scenarios (such as the standard CDM model) provide predictions which agree reasonably well with observations for galaxies (Valageas \& Schaeffer 1999a; Kauffmann et al.1993; Cole et al.1994), Lyman-$\alpha$ clouds (Valageas et al.1999a; Petitjean et al.1992; Miralda-Escude et al.1996; Riediger et al.1998), quasars and reionization constraints (Valageas \& Silk 1999a,b; Gnedin \& Ostriker 1997; Haiman \& Loeb 1998). However, within this framework the simplest model leads for clusters to a temperature - X-ray luminosity relation $L_X \propto T^2$ (Kaiser 1986) which disagrees with observations (Ponman et al.1996). It has been argued (Evrard \& Henry 1991; Cavaliere et al.1997; Ponman et al.1998) that this discrepancy could be explained by a ``preheating'' of the gas which would raise its entropy before clusters form. Indeed, this entropy ``floor'' would lead to a maximum density for the ICM which would break the previous self-similar scaling and provide a steeper relation $T-L_X$. On the other hand, it has also been suggested (Blanchard et al.1992; Blanchard \& Prunet 1997) that the overcooling problem linked to galaxy formation (i.e. the fact that a small fraction of the baryonic content of the universe has been converted into stars while simple estimates show that most of the baryons should have been able to cool by now) requires a reheating of the IGM in order to prevent the gas from cooling and falling into the dark matter potential wells. Thus, it is important to obtain a good handle on the ``entropy history'' of the universe since it may play a key role in structure formation processes, for galaxies as well as for clusters. 

In this article, we present an analytic model to derive the evolution of the 
entropy of the gas. To this order, we use a description developed in Valageas 
\& Silk (1999a,b) to study the reheating and reionization history of the 
universe by the radiation emitted by stars and quasars. This model also 
provides a consistent description of galaxies (Valageas \& Schaeffer 1999a), 
Lyman-$\alpha$ clouds (Valageas et al.1999a) and clusters (Valageas \& 
Schaeffer 1999b). Thus, it ensures that we obtain a realistic scenario. Then, as 
detailed in Sect.\ref{Model} we add to this simple model an additional source 
of heating which corresponds to a direct energy input into the IGM from 
supernovae or quasars. We also describe the modifications induced by entropy 
considerations. Next, in Sect.\ref{Numerical results} we present the numerical 
results we obtain for an open universe for both supernova and quasar heating. 
Finally, in Sect.\ref{Feedback on structure formation} we point out the 
feedback of such an entropy production onto structure formation, for galaxies, 
quasars and clusters. We also describe the cluster temperature - X-ray luminosity relation we obtain from these scenarios.

\section{Model}
\label{Model}

First, we briefly present the main characteristics of our model. Some of them 
are described in more details in Valageas \& Silk (1999a) (hereafter VS).

\subsection{Multiplicity functions}
\label{Multiplicity functions}

In order to evaluate the reheating and reionization of the universe by stars 
and quasars we need the mass functions of galaxies and QSOs. We also derive 
the multiplicity function of Lyman-$\alpha$ absorbers which provide most of 
the opacity at low $z$ after reionization. This is necessary in order to model 
the density fluctuations in the IGM itself. Indeed, at low $z$ we consider 
density contrasts from $(1+\Delta) \sim 10^{-3}$ (voids) up to $(1+\Delta) 
\sim 20$ (massive Lyman-$\alpha$ forest clouds) within the IGM. Of course, the 
need to 
describe such a large class of objects, defined by {\it various} density 
thresholds (which will also be required by entropy considerations, as shown 
below) or other constraints (e.g. on their size), means that we cannot use the 
familiar Press-Schechter prescription (Press \& Schechter 1974). Indeed, the 
latter is restricted to ``just-virialized'' halos defined by a density contrast 
$\Delta_c \sim 177$. Thus, we assume that the non-linear density field is well 
described by the scaling model developed in Balian \& Schaeffer (1989). This 
description is based on the assumption that the many-body correlation functions 
obey specific scaling laws which can also be seen as a consequence of the 
stable-clustering ansatz (Peebles 1980):
\beq
\xi_p(\lambda {\bf r}_1,...,\lambda {\bf r}_p) = \lambda^{-\gam(p-1)} \; \xi_p({\bf r}_1,...,{\bf r}_p)  
\label{scal1}
\eeq
where $\gam$ is the local slope of the two-point correlation function. This model has been checked through the counts-in-cells statistics against various numerical simulations (e.g. Colombi et al.1997; Valageas et al.1999b; Munshi et al.1999). Then, for a given class of objects defined by a relation $\Delta(M,z)$ (which implies a specific radius $R(M,z)$) we attach to each object the parameter $x$ defined by:
\beq
x(M,z) =  \frac{1+\Delta(M,z)}{\; \xia[R(M,z),z] \;}
\label{xnl}
\eeq
where 
\[
\xia(R) =   \int_V \frac{d^3r_1 \; d^3r_2}{V^2} \; \xi_2 ({\bf r}_1,{\bf r}_2) 
 \;\;\;\;\; \mbox{with} \;\;\;\;\; V= \frac{4}{3} \pi R^3
\]
is the average of the two-body correlation function $\xi_2 ({\bf
r}_1,{\bf r}_2)$ over a spherical cell of radius $R$ and provides the
measure of the density fluctuations in such a cell. Then, we write the
multiplicity function of these objects (defined by the constraint
$\Delta(M,z)$) as (Valageas \& Schaeffer 1997):
\beq
\eta(M,z) \frac{dM}{M}  = \frac{\rhob}{M} \; x^2 H(x) \; \frac{dx}{x}    
\label{etah}
\eeq
where $\rhob$ is the mean density of the universe at redshift $z$,
while the mass fraction in halos of mass between $M$ and $M+dM$ is:
\beq
\mu(M,z) \frac{dM}{M} = x^2 H(x) \; \frac{dx}{x}     
\label{muh}
\eeq
The scaling function $H(x)$ only depends on the initial spectrum of the density 
fluctuations and must be obtained from numerical simulations. In practice, we 
use the scaling function $H(x)$ obtained by Bouchet et al.(1991) for a
CDM universe. The mass functions obtained in this way have been checked 
against the results of numerical simulations for various constraints 
$\Delta(M)$ in the case of a critical universe with an 
initial power-law power-spectrum (Valageas et al.1999b). This description also takes into account the 
substructures which may exist within larger objects and it allows one to 
derive for instance the amplitude of the density fluctuations in the IGM (i.e. 
the ``clumping factor''). On the other hand, note that a simpler model where 
the density field is described as a collection of {\it smooth} halos with a 
universal density profile is inconsistent with the results of numerical 
simulations, as shown in Valageas (1999) (it implies a wrong behaviour of the 
many-body correlation functions).

\subsection{Galaxies}
\label{Galaxies}

Along the lines of VS, we use a simplified version of the model described in Valageas \& Schaeffer (1999a). We define galaxies by two constraints: 1) {\it a virialization condition} $\Delta>\Delta_c(z)$ (where $\Delta_c \sim 177$ is given by the usual spherical model) and 2) {\it a cooling condition} $t_{cool} < t_H$ which states that the gas must have been able to cool within a few Hubble times $t_H$ {\it at formation} in order to fall into the dark matter potential well and form a galaxy (see also Rees \& Ostriker 1977; Silk 1977). At late times, the second condition 2) is the most restrictive for just-virialized halos (defined by $\Delta=\Delta_c(z)$) with a large virial temperature $T$. This means that these objects are groups or clusters which contain several subunits (i.e. galaxies) which satisfy the constraint 2) which approximately translates in this case into a constant ``cooling radius'' $R_{cool} \sim 100$ kpc (see Valageas \& Schaeffer 1999a for a detailed discussion). Thus, we define galaxies by the relation $\Delta_{gal}(M,z)$ such that $\Delta_{gal}=\Delta_c(z)$ except for high temperature halos at low $z$ where this constraint would imply $R>R_{cool}$. Then, these objects are defined by the condition $R=R_{cool}$ which gives their density contrast $\Delta_{gal}(M,z)$. This allows us to distinguish clusters or groups from galaxies. In practice, as seen in Valageas \& Schaeffer (1999a), the condition 2) only plays a role at low redshift $z<1$ for high temperature galaxies ($T> 10^6$ K). Finally, we also require the virial temperature $T$ of the galactic halos to be larger than a ``cooling temperature'' $T_{cool}(z)$. The latter corresponds to the smallest just-virialized objects which can cool efficiently at redshift $z$, defined by the constraint $t_{cool} = t_H$. Moreover, it is constrained to be larger than or equal to the temperature of the IGM. At low $z$ we have $T_{cool} \sim 3\;10^4$ K in our original model (VS) since for smaller temperatures cooling is very inefficient (due to recombination). From the lower bound $T_{cool}$ and the density threshold $\Delta_{gal}(M,z)$ we obtain the galaxy mass function using (\ref{etah}).

Then, we use a simple star formation model to derive the stellar content and the luminosity of these galaxies. This involves 4 components: (1) short lived stars which are recycled, (2) long lived stars which are not recycled, (3) a central gaseous component which is deplenished by star formation and ejection by supernovae winds, replenished by infall from (4) a diffuse gaseous component. The star formation rate $dM_s/dt$ is proportional to the mass of central gas with a time-scale set by the dynamical time. The mass of gas ejected by supernovae is proportional to the star formation rate and decreases for deep potential wells as $1/T$, in a fashion similar to Kauffmann et al.(1993). Some predictions of this model (galaxy luminosity function, Tully-Fisher relation) have already been checked against observations (Valageas \& Schaeffer 1999a). Thus, we obtain for the mean star-formation rate per Mpc$^3$:
\beq
\left( \frac{d\rho_s}{dt} \right) = \frac{\Omega_b}{\Omega_m} \; \frac{\rhob(z)}{t_H} \; \int_{x_{cool}}^{\infty} \lambda(x) \; e^{-\lambda(x)} \; x^2 H(x) \; \frac{dx}{x}  
\label{SFRav}
\eeq
with:
\beq
\lambda(x) = \frac{p}{\beta_d} \left(1+\frac{T_{SN}}{T} \right)^{-1} \sqrt{ \frac{(1+\Delta)_{gal}(x)}{(1+\Delta_c)} }
\label{lambdax}
\eeq
where $p/\beta_d = 0.5$ is a parameter of order unity which enters the definition of the dynamical time, while $T_{SN} = 10^6$ K describes the ejection of gas by supernovae and stellar winds (see also Kauffmann et al.1993):
\beq
T_{SN} =  \frac{2 \; \epsilon_{SN} \; E_{SN} \; \mu m_p \; \eta_{SN}}{3 \; k \; m_{SN}} = 10^{6} \; \mbox{K}  
\label{TSN}
\eeq
Here $\mu$ is the mean molecular weight, $\epsilon_{SN} \sim 0.1$ is the 
fraction of the energy $E_{SN}$ delivered by supernovae transmitted to the 
gas ($E_{SN} = 10^{51}$ erg) while $\eta_{SN}/m_{SN} \simeq 0.005 \; 
M_{\odot}^{-1}$ is the number of supernovae per solar mass of stars formed 
(note that in VS we used $T_{SN}=2 \; 10^6$ K). 
The factor $\lambda \; e^{-\lambda}$ in (\ref{SFRav}) comes from the dependance 
of the efficiency of star formation on the properties of the host galaxy. Thus, 
small galaxies with a shallow potential well ($T \ll T_{SN}$) are strongly 
influenced by supernovae and stellar winds which eject part of the gas so that 
a small fraction of the baryonic matter is converted into stars ($\lambda \ll 
1$). On the other hand, at low $z$ large halos ($T \gg T_{SN}$) which formed at 
a high redshift ($\Delta_{gal} > \Delta_c$) have already converted most of 
their gas into stars ($\lambda \gg 1$). Indeed, their high density (due to 
their large redshift of formation) translates into a small dynamical time, 
hence to very efficient star formation within our model, which leads to the 
factor $\sqrt{(1+\Delta)_{gal}}$ in (\ref{lambdax}). As in the model used in 
Valageas \& Schaeffer (1999a) and VS we assume that the 
gas ejected from the inner parts of small galaxies ($T \ll T_{SN}$) remains 
bound to (or close to) the galactic halo so that it can cool and fall back into 
the galaxy. This leads to a self-regulated star formation process (within each 
individual galaxy) which takes care of the ``overcooling problem''. Note that 
an alternative, as suggested in Blanchard et al.(1992) (also Prunet \& Blanchard 1999), would be that this gas 
gets mixed with the IGM and leads to a progressive heating of the IGM which 
prevents most of the gas to cool and form galaxies. This will also correspond 
to our supernova heating scenario (SN), see Sect.\ref{Evolution of the IGM}.

\subsection{Quasars}
\label{Quasars}

In a fashion similar to Efstathiou \& Rees (1988) and Nusser \& Silk (1993) we derive the quasar luminosity function from the multiplicity function of galactic halos. Thus, we assume that the quasar mass $M_Q$ is proportional to the mass of gas $M_{gc}$ available in the inner parts of the galaxy: $M_Q = F \; M_{gc}$. In our case this also implies that $M_Q \simeq F \; M_s$ where $M_s$ is the stellar mass. We use $F=0.008$ which is consistent with observations (Magorrian et al.1998 find that $M_Q \sim 0.006 \; M_s$). We also assume that quasars shine at the Eddington limit so that their life-time is given by $t_Q = 4.4 \; \epsilon_Q \; 10^8$ yr where $\epsilon_Q = 0.1$ is the quasar radiative efficiency and that a fraction $\lambda_Q=0.1$ of galactic halos actually host a quasar. Thus, the luminosity of a quasar of mass $M_Q$ is:
\beq
L_Q = \frac{\epsilon_Q \; M_Q \; c^2}{t_Q} 
\label{LQ}
\eeq 
and the quasar multiplicity function $\eta_Q(M_Q) dM_Q/M_Q$ is obtained from the galaxy mass function $\eta_g(M) dM/M$ by:
\beq
\eta_Q(M_Q) \frac{dM_Q}{M_Q} = \mbox{Min} \left[ 1 , \frac{t_Q}{t_M} \right] \; \lambda_Q \; \eta_g(M) \frac{dM}{M} 
\label{etaQ}
\eeq
Here $t_M$ is the evolution time-scale of galactic halos of mass $M$ defined by:
\beq
t_M^{-1} = \frac{1}{\eta_g(M)} \; \frac{\pl}{\pl t} \eta_g(M)
\eeq
Since the quasar life-time $t_Q \sim 10^8$ yr is quite short, we have 
$\eta_Q(M_Q) dM_Q/M_Q = \lambda_Q \; t_Q \; \pl \eta_g / \pl t \; dM/M$. This
also means that $\eta_Q(M_Q) dM_Q/M_Q \sim \lambda_Q \; t_Q/t_H \; \eta_g 
\; dM/M$. The factor $t_Q/t_H$ shows that the quasar luminosity function
is biased towards large redshifts as compared with the galaxy luminosity 
function. In particular, it peaks at $z \sim 2$ and shows a significant drop 
at smaller redshift while the galaxy luminosity function keeps increasing until
$z \sim 0$ and the star formation rate only decreases after $z \la 1$. As we 
shall see in Sect.\ref{Numerical results} this implies that quasar heating
of the IGM occurs earlier than supernova heating. Note that 
we only have two parameters: $(\epsilon_Q \; F/t_Q)$ and $(\lambda_Q \; t_Q)$. 
Hence a larger fraction of quasars $\lambda_Q$ with a smaller life-time $t_Q$ 
would give the same results. Moreover, the assumption that quasars shine at 
the Eddington limit gives the ratio $(\epsilon_Q/t_Q)$ while the parameter $F$ 
is constrained by the observed ratio (quasar mass)/(stellar mass). The 
normalization factor $(\lambda_Q \; t_Q)$ is constrained by the observed quasar 
luminosity function. Our results agree reasonably well with available B-band 
observations for $0.16<z<4.5$ (VS and Sect.\ref{Quasar 
luminosity function}).

\subsection{Lyman-$\alpha$ clouds}
\label{Lyman clouds}

We also include in our model a description of Lyman-$\alpha$ clouds. These correspond to density fluctuations in the IGM as well as to virialized halos which may or may not have cooled. More precisely, we consider three different classes of objects (Valageas et al.1999a). 

Low-density mass condensations with a small virial temperature see their baryonic density fluctuations erased over a scale $R_d$ as the gas is heated by the UV background radiation (or other processes) to a temperature $T_{forest}$. More precisely, we define the scale $R_d$ by:
\beq 
R_d(z) = \frac{1}{2} \; t_H \; C_s = \frac{1}{2} \; t_H \; \sqrt{ \frac{\gam k T_{IGM}}{\mu m_p} }
\label{Rd}
\eeq 
where $C_s$ is the sound speed, $T_{IGM}$ the IGM temperature, $t_H$ the age of the universe, $m_p$ the proton mass and $\gam \sim 5/3$. These mass condensations form a first population of objects, defined by the scale $R_d$, which can be identified with the Lyman-$\alpha$ forest at low $z$. We set the characteristic temperature $T_{forest}$ by:
\beq
\left\{ \begin{array}{l} {\displaystyle  z<z_{ri} \; : \; T_{forest}= \mbox{Max} \left( 3 \; 10^4 \mbox{K} , T_{IGM} \right) } \\ \\ {\displaystyle z>z_{ri} \; : \; T_{forest}= T_{IGM} } \end{array} \right.
\label{Tforest}
\eeq
where $z_{ri}$ is the reionization redshift and $T_{IGM}$ is the temperature of the IGM. At low $z$ the term $3 \; 10^4$ K models photoionization heating for the clouds (the IGM is also heated by the UV flux but in addition it undergoes adiabatic cooling because of the expansion of the universe, so that at low redshift $z<1$ we can have $T_{IGM} < 3 \; 10^4$ K). As explained in Valageas et al.(1999a), these absorbers are not necessarily spherical clouds of radius $R_d$. Some may be long filaments with a length $L \gg R_d$ and a thickness $R_d$. Moreover, they can also be interpreted as density fluctuations within the IGM rather than distinct entities. Next, potential wells with a larger virial temperature $T > T_{forest}$ do not see their baryonic density profile smoothed out. Thus, they define a second class of absorbers which for $T > T_{cool}$ correspond to the galactic halos (objects with $T_{forest} < T < T_{cool}$, if such a range exists, simply are virialized objects which have not cooled hence have not formed stars). Note that one such object can produce a broad range of observed column densities depending on the impact parameter of the line of sight. This population corresponds to Lyman-limit systems. Finally, the deep cores of these halos are neutral because of self-shielding and they form our third class of absorbers, corresponding to damped systems.

\subsection{Evolution of the IGM}
\label{Evolution of the IGM}

\subsubsection{Temperature evolution}
\label{Temperature evolution}

As in VS the gas in the IGM is heated by the background radiation while it cools because of the expansion of the universe and several atomic processes (collisional excitation, ionization, recombination, molecular hydrogen cooling, bremsstrahlung and Compton cooling or heating). Meanwhile, hydrogen and helium are reionized by the UV flux. In our calculation, we take into account the opacity due to the gas present in the underdense regions which fill most of the volume as well as the absorption due to discrete clouds (the ``Lyman-$\alpha$ clouds'' described above). We also follow the evolution of the HI, HeII and HeIII filling factors describing the ionized bubbles around galaxies and quasars, as well as the clumping of the gas (which also enters explicitly into the model for Lyman-$\alpha$ clouds). The evolution of the background radiation field $J_{\nu}$ is obtained from the radiation emitted by stars and quasars, which we described in the previous sections, see VS for details. We write the evolution of the temperature of the IGM as:
\beq 
\frac{dT_{IGM}}{dt} = - 2 \; \frac{\dot{a}}{a} \; T_{IGM} \; - \; \frac{T_{IGM}}{t_{cool}} \; + \; \frac{T_{IGM}}{t_{heat,J}} \; + \; \frac{T_{IGM}}{t_E}
\label{TIGM} 
\eeq 
where $a(t)$ is the scale factor (which enters the term describing adiabatic cooling due to the expansion). The heating time-scale $t_{heat,J}$ which corresponds to photoionization heating is given by: 
\beq
t_{heat,J}^{-1} = \frac{4 \pi}{3/2 n_b k T_{IGM}} \; \sum_j \; \int n_j
\sigma_j(\nu) (\nu - \nu_j) J_{\nu} \frac{d\nu}{\nu} 
\label{theat}
\eeq 
where $j=$ (HI,HeI,HeII), $\nu_j$ is the ionization threshold of the corresponding species, $n_j$ its number density in the IGM and $n_b$ the baryon number density. The cooling time-scale $t_{cool}$ describes collisional excitation, collisional ionization, recombination, molecular hydrogen cooling, bremsstrahlung and Compton cooling or heating. We compute the redshift evolution of the ionization state of hydrogen and helium and we use the cooling rates from Anninos et al.(1997). In particular, as shown in the upper panel of Fig.4 in VS at low $z$ after reionization the main cooling processes in the IGM are adiabatic and Compton cooling. Indeed, when the medium is reionized collisional excitation cooling is strongly suppressed (see discussion in VS and Efstathiou 1992).

Finally, we added to the evolution equation we used in VS a new term $T_{IGM}/t_E$. This corresponds to an additional source of energy, which we assume here to be uniform. In particular, this term models in our framework the energy output provided by supernovae or quasars, which has been advocated in the litterature in order to raise the entropy level of the IGM (e.g. Ponman et al.1998; Tozzi \& Norman 1999). As explained in Sect.\ref{Galaxies}, in our original model the influence of supernovae was restricted to their parent galaxy (note that this is consistent with numerical simulations by Mac Low \& Ferrara 1999 which suggest that gas ejection is negligible for galactic halos with $M>10^6 M_{\odot}$). In contrast, one model we investigate in this article corresponds to a ``maximally efficient'' scenario where the energy produced by supernovae reheats the IGM as a whole. In the actual universe, the effect of supernovae is likely to lie somewhere in-between these two cases, but these two models allow us to get an estimate of the allowed range for the reheating process (see also Tegmark et al.1993 for a study of reheating and reionization of the IGM by supernovae-driven winds). 

We note that using a uniform source of energy (i.e. we do not let the energy source term vary in space as a function of the distance to the nearest galaxy or quasar, although we model this effect for photoionization heating) is probably a better approximation than it may seem at first sight. Indeed, at late times $z \la 2$ when this process dominates most of the matter is embedded within positive density fluctuations (filaments, virialized halos, see VS) which show a strong clustering pattern as seen in numerical simulations (e.g. Bond et al.1996). Note that this is included in our model of the density field, described in Sect.\ref{Multiplicity functions}. For instance, we obtained in Valageas et al.(1999a) the amplitude of the two-point correlation function of Lyman-$\alpha$ clouds and we describe in Valageas et al.(1999c) the bias of the various objects we observe in the universe (Lyman-$\alpha$ clouds, galaxies, quasars, clusters). Thus, most of the volume consists of low-density regions while most of the matter is embedded within small or thin structures (filaments, halos) which are located close to galaxies since most clusters and galaxies form on density peaks within these mass condensations, though there may also be some isolated galaxies amid low-density regions (note that this ``bias'' translates into the correlations of these objects). As a consequence, the energy provided by supernovae or quasars does not need to travel very far in order to heat most of the matter. Indeed, for this it is sufficient to ``spread'' the energy over filaments while leaving cool voids in between. In this case, the temperature $T_{IGM}$ would rather correspond to a ``mass-averaged'' temperature, describing the network of halos and filaments which contain most of the matter while voids would be cooler. Of course, one may also expect voids to be easily heated to the temperature of the filaments since due to their low density and small mass they only require a small amount of energy in order to reach the temperature of the neighbouring regions. Thus, the assumption of a uniform energy source appears to be a reasonable first order approximation. However, it is clear that a carefull study of this problem would be interesting, but this would probably require very detailed numerical simulations which are beyond the scope of this study.

In this article, we consider the additional energy described by the term $T_{IGM}/t_E$ in (\ref{TIGM}) to be provided by supernovae or quasars. Thus, we can write:
\beq
\frac{T_{IGM}}{t_E} = \frac{T_{IGM}}{t_{E,SN}} + \frac{T_{IGM}}{t_{E,Q}}
\label{tE}
\eeq
which explicitly shows these two possible sources of energy. Using our model for galaxies which we described in Sect.\ref{Galaxies}, we can write the source term $T_{IGM}/t_{E,SN}$ due to supernovae as:
\beq
\frac{T_{IGM}}{t_{E,SN}} = \frac{\alpha_{SN}}{0.1} \; \frac{T_{SN}}{\rhob_b(z)} \; \left( \frac{d\rho_s}{dt} \right) 
\label{tSN}
\eeq
where $\rhob_b(z)= \Omega_b/\Omega_m \; \rhob(z)$ is the mean baryonic density 
of the universe (the fraction of matter within stars is always negligible) and 
$\alpha_{SN}$ is the efficiency factor, similar to $\epsilon_{SN}$ in 
(\ref{TSN}), which measures the fraction of the energy produced by supernovae 
which is available to heat the gas. Thus, we have $\alpha_{SN} \leq 1$. Next, from 
the model of quasars presented in Sect.\ref{Quasars} we have for the quasar 
contribution:
\beq
\frac{T_{IGM}}{t_{E,Q}} = \frac{\alpha_Q}{0.1} \; \frac{1}{3/2 n_b k} \; \int L_Q \; \eta_Q(L_Q) \frac{dL_Q}{L_Q}
\label{tQ}
\eeq
where $\alpha_Q \leq 1$ is the efficiency factor similar to $\epsilon_Q$ in (\ref{LQ}). However, if there are some additional energy sources (e.g. the decay of some exotic particles) we could have an effective $\alpha$ larger than unity. Of course, in this case the time-dependence of this hypothetic energy source is unlikely to be proportional to the star or quasar formation rate and one should explicitly detail the origin of this process to get its time-evolution. In this article, we shall restrict ourselves to the formulation (\ref{tE}) which models the possible effect of supernovae or quasars on the IGM, but one cannot disregard the fact that our source term $T_{IGM}/t_E$ may in fact correspond to some new process. From the expressions (\ref{tSN}) and (\ref{tQ}) we can directly obtain a simple estimate of the magnitude of these effects. Indeed, from (\ref{tSN}) we see that supernovae heat the IGM to a temperature $T_{IGM,SN}$ of the order:
\beq
T_{IGM,SN} \sim \frac{\alpha_{SN}}{0.1} \; F_{star} \; T_{SN} \sim \alpha_{SN} \; 10^6 \; \mbox{K}
\label{TIGMSN}
\eeq
where we used (\ref{TSN}) and the fact that at late times $z<1$ the fraction of baryonic matter which has been converted into stars is $F_{star} \sim 0.1$. Thus, {\it supernovae can reheat the IGM up to $10^6$} K {\it at most}. On the other hand, from (\ref{tQ}), (\ref{LQ}) and (\ref{etaQ}) we see that quasars heat the IGM up to $T_{IGM,Q}$ of the order:
\beq
T_{IGM,Q} \sim \frac{\alpha_Q}{0.1} \; (\lambda_Q t_Q) \; \left( \frac{\epsilon_Q F}{t_Q} \right) \; F_{star} \; T_Q \sim \alpha_Q \; 10^8 \; \mbox{K}
\label{TIGMQ}
\eeq
where we used the parameters introduced in Sect.\ref{Quasars} and we defined:
\beq
T_Q = \frac{2 \mu m_p c^2}{3 k} \simeq 4.3 \; 10^{12} \; \mbox{K}
\label{TQ}
\eeq
The factor $F_{star}$ comes from the fact that $M_Q = F \; M_{gc} \sim F \; 
M_s$, which agrees with observations. Thus, {\it quasars can potentially heat the 
IGM to a very high temperature}, much larger than the temperature induced by 
supernova heating, because quasars are very efficient engines to convert the 
rest mass energy of matter into radiation or energy while a small fraction of 
the matter converted into stars leads to supernovae ($\sim 10^{-3}$) which 
themselves have a small efficiency factor ($\sim 10^{-3}$) so that $T_{SN} 
\sim 10^{-6} T_Q$. Note that the estimates (\ref{TIGMSN}) and (\ref{TIGMQ}) 
are very robust, independently of the details of the model, since they are 
directly constrained by the observed galaxy and quasar luminosity functions. 
On the other hand, the new parameters $\alpha_{SN}$ and $\alpha_Q$ are only 
constrained to be smaller than unity. One would need a detailed study of many 
physical processes which are still poorly known to set a precise value for 
these efficiency factors. In this article, we shall treat them as free 
parameters, which we take to be constant in time. Thus, our goal is to evaluate 
the possible effects of these energy sources, which in turn will give us some 
constraints on their magnitude.

\subsubsection{Entropy evolution}
\label{Entropy evolution}

From the model of the IGM described in the previous sections, we can also obtain the evolution of the entropy of the gas. The entropy of a Maxwell-Boltzmann gas is given by the Sackur-Tetrode equation:
\beq
S_{M.B.}(N,V,T) = N k \left[ \ln \left( \frac{V}{N \lambda_T^3} \right) + \frac{5}{2} \right]
\label{SMB}
\eeq
where $N$ is the number of particles, within the volume $V$, and:
\beq
\lambda_T = \sqrt{ \frac{2\pi\hbar^2}{m k T} }
\eeq
Thus we define the specific entropy $S$ as:
\beq
S = \log \left( \frac{k T \; n_b^{-2/3}}{1 \; \mbox{keV cm}^2} \right)
\label{S}
\eeq
where $n_b$ is the baryonic number density (and we note $\log$ the decimal logarithm). As explained in details in VS, we consider that at late times most of the volume of the IGM consists of large underdense regions with a density contrast $(1+\Delta)_u$ (``u'' for underdense) given by:
\beq
(1+\Delta)_u = \mbox{Min} \left[ \;1 \; , \; \xia(R_d)^{-\omega/(1-\omega)} \; \right]
\label{Deltau}
\eeq 
This simply states that at high $z$ (when $\xia(R_d) \ll 1$) we have $\rho_u = \rhoa$ (i.e. the universe is almost exactly a uniform medium on scale $R_d$) while at low $z$ we have $\rho_u < \rhoa$ since most of the matter is now within overdense objects (clusters, filaments, etc.) while most of the volume is formed by underdense regions. The scale $R_d$ was defined in (\ref{Rd}) while the exponent $\omega$ is given by the power-law behaviour of the scaling function $H(x)$ at small $x$. In addition to these ``voids'' and the virialized halos which are identified to galaxies or clusters, there are also density fluctuations (clouds, filaments) which form the Lyman-$\alpha$ forest or small virialized halos which have not cooled ($T<T_{cool}$) and can have a larger temperature than the underdense regions (the temperature of the gas within these small halos is of the order of the virial temperature of the potential well). It is important to take into account these density fluctuations since at late times ($z<1$) the ``overdensity'' $(1+\Delta)_u$ is as low as $\sim 10^{-2}$ while the density contrast of forest clouds reaches $\sim 20$. This wide variation of the local density means that the average entropy of the IGM can be significantly different from the entropy which would be computed from (\ref{S}) within ``voids''. Thus, we first define the entropy $S_u$ characteristic of the underdense regions which fill most of the volume by:
\beq
S_u = \log \left( \frac{k T_{IGM} \; n_u^{-2/3}}{1 \; \mbox{keV cm}^2} \right)
\label{Su}
\eeq
where $n_u$ is the baryonic density obtained from (\ref{Deltau}). Then, we define a ``mean IGM entropy'' $\lag S \rag_{IGM}$ by:
\beq
\begin{array}{l} {\displaystyle \lag S \rag_{IGM} = \frac{1}{\lag 1+\Delta \rag_{IGM}} \; \Biggl \lbrace (1+\Delta)_u \; S_u } \\ \\ {\displaystyle \hspace{3cm} + \int_0^{x_{cool}} S(x) \; x^2 H(x) \frac{dx}{x} \Biggl \rbrace } \end{array}
\label{SIGM}
\eeq
where the mean density contrast $\lag 1+\Delta \rag_{IGM}$ is given by:
\beq
\lag 1+\Delta \rag_{IGM} = (1+\Delta)_u + \int_0^{x_{cool}} x^2 H(x) \frac{dx}{x}
\label{DeltaIGM}
\eeq
It corresponds to the mean density (total mass over the volume) of the matter which is not embedded within virialized halos which have cooled, from very underdense regions up to forest clouds seen as density fluctuations in the IGM. Since the entropy is an additive quantity, the relevant quantity is indeed the mean $\lag S \rag_{IGM}$ which describes the average entropy of the IGM, see (\ref{SMB}). In particular, at late times the quantity $S_u$ which corresponds to the small fraction of matter located within voids is much larger. We also define the mass-averaged overdensity $\lag 1+\Delta \rag_{Ly}$ characteristic of the matter outside galaxies and clusters by:
\beq
\begin{array}{l} {\displaystyle \lag 1+\Delta \rag_{Ly} = \frac{1}{\lag 1+\Delta \rag_{IGM}} \; \Biggl \lbrace (1+\Delta)_u^2 } \\ \\ {\displaystyle \hspace{2cm} + \int_0^{x_{cool}} (1+\Delta)(x) \; x^2 H(x) \frac{dx}{x} \Biggl \rbrace } \end{array}
\label{DeltaLy}
\eeq
The quantity $\lag 1+\Delta \rag_{Ly}$ corresponds to the average overdensity of IGM particles, weighted by the number of particles and not by the volume they occupy. Finally we introduce the mean temperature $T_{Ly}$:
\beq
\begin{array}{l} {\displaystyle T_{Ly} = \frac{1}{\lag 1+\Delta \rag_{IGM}} \; \Biggl \lbrace (1+\Delta)_u \; T_{IGM} } \\ \\ {\displaystyle \hspace{2.5cm} + \int_0^{x_{cool}} T(x) \; x^2 H(x) \frac{dx}{x} \Biggl \rbrace } \end{array}
\label{TLy}
\eeq
This takes into account the fact that some of the gas outside galaxies and clusters is located in Lyman-$\alpha$ forest clouds with $T>3 \; 10^4$ K (due to photoionization heating) which may be larger than $T_{IGM}$ (which also involves adiabatic cooling), and possibly within some virialized halos with $T_{IGM} < T < T_{cool}$ which have not cooled (due to their low virial temperature which implies inefficient cooling).

\subsubsection{Effect of the IGM entropy on galaxy formation}
\label{Effect of the IGM entropy on galaxy formation}

As gravitational clustering builds increasingly large structures, the baryonic matter content of the universe gradually becomes embedded into virialized halos where it cools and forms stars, as described in Sect.\ref{Galaxies} where we detailed our model for galaxy formation. In particular, the ``cooling temperature'' $T_{cool}(z)$ which characterizes the smallest virialized halos which can cool was given by the condition $t_{cool}=t_H$ where the cooling time which depends on the density and the temperature of the gas satisfies:
\beq
t_{cool} \propto \frac{T}{n_b \Lambda(T)}
\label{tcool}
\eeq
where $\Lambda(T)$ is the cooling function. Thus cooling is more efficient for larger baryonic densities 
$n_b$ (because collisions are more frequent). In the original model the cooling time attached to a given 
halo was computed using the virial temperature $T_{vir}$ for $T$ and a density contrast $\Delta_c(z) \sim 177$ 
to obtain the gas density. However, if the IGM is preheated and gets a large entropy at earlier times, the gas 
may not follow the dark matter to form a mass condensation with mean baryonic density $\rho_b = (1+\Delta_c) 
\rhob_b$. Indeed, during the adiabatic collapse of the gas (before it cools) its temperature increases as
 $T_{ad} \propto n_b^{2/3}$. Hence the compression will stop if $T_{ad}$ reaches $T_{vir}$ before the density 
 contrast reaches $\Delta_c$. Indeed, gas with $T>T_{vir}$ does not fall into the potential well. Thus, we obtain 
 an upper bound $n_{b,ad}$ for the baryonic density reached within a virialized halo of temperature $T_{vir}$, 
 defined by:
\beq
\lag S \rag_{IGM} = \log \left( \frac{k T_{vir} \; n_{b,ad}^{-2/3}}{1 \; \mbox{keV cm}^2} \right)
\label{nbad}
\eeq
We can also write the density contrast $(1+\Delta)_{ad}$ given by (\ref{nbad}) as:
\beq
(1+\Delta)_{ad} = (1+\Delta)_u \; \left( \frac{T_{vir}}{T_{IGM}} \right)^{3/2} \; 10^{ \frac{3}{2} ( S_u - \lag S \rag_{IGM} ) }
\label{Deltaad}
\eeq
Thus, in order to compute the cooling time $t_{cool}$ from (\ref{tcool}) we use the overdensity $(1+\Delta)_b$ given by:
\beq
\Delta_b = \mbox{Min} \left( \Delta_c , \Delta_{ad} \right)
\label{Deltab}
\eeq
Using this prescription we compute the ``cooling temperature'' $T_{cool}(z)$. 
It is obvious from (\ref{tcool}) and (\ref{Deltab}) that {\it the effect of the 
entropy of the intergalactic gas is to make cooling less efficient, which 
leads to a possible increase of the characteristic temperature $T_{cool}$}. In 
particular, at late times ($z < 3$) if the entropy production is sufficiently 
large it may happen that the cooling temperature $T_{cool}$ does not exist any 
more. Indeed, as we explained above the large entropy of the gas diminishes 
the gas density which enters (\ref{tcool}). Moreover, the ratio $T/\Lambda(T)$ 
displays a minimum at a finite value $T \sim 10^5$ K since at large 
temperatures the cooling function behaves as $\Lambda(T) \propto \sqrt{T}$ 
(bremsstrahlung is the main cooling process) while below $10^4$ K it nearly 
goes to 0 (note that in our original model the high temperature halos which 
cannot cool are identified to clusters while the low temperature objects are 
Lyman-$\alpha$ absorbers). As a consequence, if the entropy level is such that 
the gas density within the halos with a virial temperature $\sim 10^5$ K is too 
small to allow efficient cooling, no halo can cool. Of course, this does not 
mean that there are no galaxies ! It simply means that all just-virialized 
halos (i.e. overdensities defined by the density threshold $\Delta_c$) are 
identified to ``clusters'' (or ``groups'') in the sense that they consist of 
one or several smaller higher-density subunits (galaxies), which could cool at 
earlier 
times when they formed, embedded within a larger structure containing some hot 
gas. 

Let us note $z_S$ the largest redshift where no halo can cool (i.e. $T_{cool}(z_S)$ as defined above does not exist). We shall have $z_S < 3$ since the entropy production is linked to galaxy or quasar formation and cooling is less efficient at low redshift where the baryonic density is lower. Moreover, since the mean entropy $\lag S \rag_{IGM}$ increases with time, at smaller redshifts $z<z_S$ we have the same situation. Then, along the lines developed in Valageas \& Schaeffer (1999a) to distinguish galaxies from clusters at $z \sim 0$, we assume that gravitational clustering is stable and that, to a first order approximation, galaxies which form at $z_S$ do not evolve significantly at later times when they get embedded within larger structures which cannot cool as a whole. Thus, after $z_S$ these galaxies may get closer to form a group but we neglect their possible mergings. Note that the gas which cooled before $z_S$ and fell into these potential wells to build these galaxies formed ``small'' dense entities (the baryonic distribution extends to smaller radii than the underlying dark matter halo) which are likely to keep their identity for a longer time than their surrounding dark matter halos which may join to make a larger object. Furthermore, we note that the presence of substructures within dark matter halos themselves and the dependance of the characteristic density of a halo on its mass (it is proportional to the average density of the universe at the time this mass-scale turned non-linear, e.g. Navarro et al.1996) suggest that this picture may also be a good approximation for the dark matter density fluctuations themselves (see discussion in Valageas 1999). Thus, we assume that after $z_S$ these galaxies keep their mass, radius and density unchanged. As a consequence, their density contrast at a later time is not $\Delta_c(z)$ but:
\beq
z < z_S \; : \; (1+\Delta)_{gal}(z) = (1+\Delta_c)(z_S) \; \left( \frac{1+z_S}{1+z} \right)^3
\label{Deltagal}
\eeq
Thus, at small redshifts $z<z_S$ we no longer define galaxies by the virialization condition $\Delta=\Delta_c(z)$. Instead, we use the constraint $\Delta=\Delta_{gal}(z)$ as defined in (\ref{Deltagal}). As explained in Sect.\ref{Multiplicity functions} this can be done in a straightforward fashion within our description of the non-linear density field: we simply use this density contrast $\Delta_{gal}(z)$ in (\ref{xnl}) to obtain the multiplicity function. Of course, as can be checked in (\ref{xnl}) and (\ref{muh}) the parameter $x$ attached to such galactic halos does not evolve with time which also implies that the fraction of matter embedded within these objects is constant with time. Thus our prescription is self-consistent. This relies on the fact that in the non-linear regime relevant for galaxies the two-point correlation function grows as $\xia(R,t) \propto a(t)^3$ at fixed physical length $R$, where $a(t)$ is the scale-factor, as predicted by the stable-clustering ansatz (Peebles 1980). This behaviour is indeed consistent with numerical simulations (e.g. Valageas et al.1999b). Then, at low redshifts we define $T_{cool}$ as being equal to the value it had at $z_S$ (more exactly for $z \rightarrow z_S^+$), with the constraint that it is larger than $T_{IGM}$:
\beq
z < z_S \; : \; T_{cool}(z) = \mbox{Max} \left[ T_{cool}(z_S) , T_{IGM} \right]
\label{TcoolzS}
\eeq
This is indeed the virial temperature of the smallest galaxies, which cooled 
at $z_S$, with $T>T_{IGM}$. This latter constraint is due to the fact that the 
gas within small halos with $T<T_{IGM}$ will be heated up to $T_{IGM}$ and it 
will escape from the potential well (but of course there will remain a small 
galaxy made of old stars). However, this condition does not play any role 
in practice since $T_{IGM}$ does not increase much after $z_S$.

\section{Numerical results}
\label{Numerical results}

We can now use the model we described in the previous sections to obtain the entropy history of the universe, as well as a consistent description of its reheating and reionization, together with the formation of quasars, galaxies and Lyman-$\alpha$ clouds. We shall consider the case of an open universe $\Omega_m=0.3$, $\Omega_{\Lambda}=0$, with a CDM power-spectrum (Davis et al.1985), normalized to $\sigma_8=0.77$. We choose a baryonic density parameter $\Omega_b=0.03$ and $H_0=60$ km/s/Mpc.

We shall consider three cases for the ``entropy scenario''. First, for reference we present some results we obtain for $\alpha_{SN}=\alpha_Q=0$, as in VS. Then, we study both cases where only one of these two efficiency factors is non-zero. This allows us to see clearly the influence of each of these processes, as well as the magnitude of the relevant parameter $\alpha$ needed to get an appreciable effect. More precisely, in both cases we choose the value of $\alpha$ such that the mean IGM entropy $\lag S \rag_{IGM}$ defined in (\ref{SIGM}) which we obtain at $z=0$ satisfies $\lag S \rag_{IGM} \simeq 2$. Indeed, Ponman et al.(1998) find that the ``entropy'' $s=T/n_e^{2/3}$ of small cool clusters seems to depart from the expected scaling law and to converge towards a floor value $s \sim 100 h^{-1/3}$ keV cm$^2$. This provides an upper bound for the IGM entropy $\lag S \rag_{IGM}(z=0)$ since these objects have just formed and we can expect the entropy of the hot gas to increase (e.g. through shocks) rather than decrease during gravitational collapse (before it cools). Moreover, if we assume that supernovae or quasars are indeed the source of this entropy floor (since gravitational collapse effects shoud not break the expected scaling law) this gives the value of the corresponding parameter $\alpha$. In the actual universe it might happen that both sources of heating, supernovae and quasars, have the same magnitude. Then, the parameters $\alpha_{SN}$ and $\alpha_Q$ would be close to those we obtained for the individual cases. However, such a coincidence would be somewhat surprising.

\subsection{Reheating of the universe}
\label{Reheating of the universe}

\begin{figure}[htb]

\begin{picture}(230,370)

\epsfxsize=15cm
\epsfysize=14cm
\put(-5,-20){\epsfbox{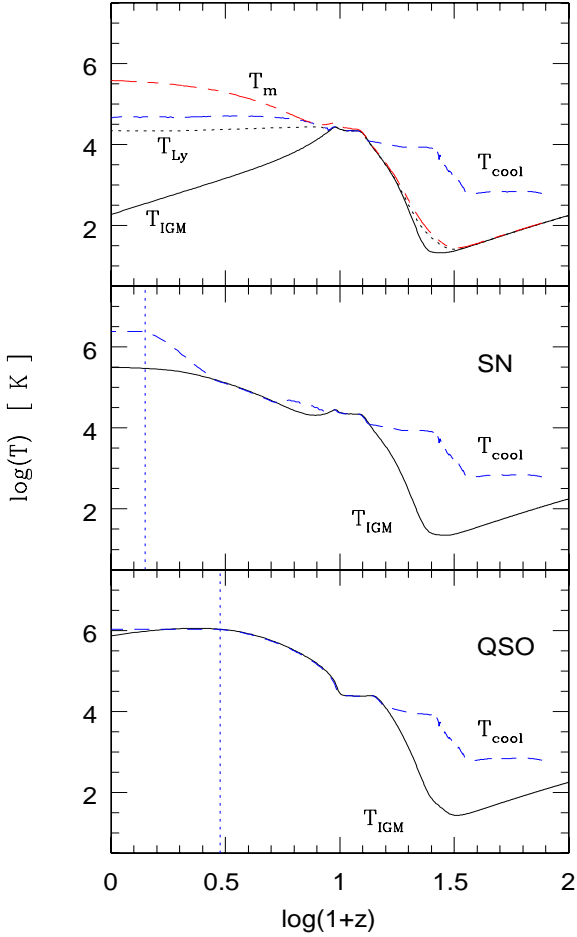}}

\end{picture}

\caption{The redshift evolution of the characteristic temperatures of the 
universe. We display the IGM temperature $T_{IGM}$ (solid curve) and the 
virial temperature $T_{cool}$ (dashed line) of the smallest galaxies. The 
upper panel corresponds to $\alpha_{SN}=\alpha_Q=0$ (only photoionization 
heating), the middle panel labelled ``SN'' to supernova heating ($\alpha_{SN} 
\neq 0$ ,  $\alpha_Q=0$) and the lower panel labelled ``QSO'' to quasar 
heating ($\alpha_{SN} = 0$ ,  $\alpha_Q \neq 0$). The vertical line in both 
lower figures shows the redshift $z_S$. In the upper panel we also show the 
mass-averaged temperature $T_m$ (dot-dashed line) and the mean temperature 
$T_{Ly}$ (dotted line) of matter located outside galaxies and clusters.}
\label{figTO03}

\end{figure}

We show in Fig.\ref{figTO03} the reheating history we obtain for the three 
cases. The temperature $T_m$ is a mass-averaged temperature which takes into 
account all the matter, from ``voids'' up to filaments, galaxies and clusters. The decrease with time of the IGM temperature at large redshift $z>24$ ($\log(1+z)>1.4$) is due to the adiabatic expansion of the universe. Then, at $z<24$ the IGM is slowly reheated by stars and quasars until it reaches at $z \sim 9$ a maximum temperature $T_{IGM} \sim 3 \; 10^4$ K where collisional excitation cooling prevents any further increase. Eventually, at low $z$ the temperature starts decreasing again because of the adiabatic expansion of the universe since the heating time becomes larger than the Hubble time (see upper panel in Fig.3 below and Sect.8.2 in VS). As described in Sect.\ref{Lyman clouds} and Valageas et al.(1999a), at low $z$ we set the characteristic temperature of Lyman-$\alpha$ clouds to $3 \; 10^4$ K when $T_{IGM}$ becomes smaller than this value. Indeed, although the density contrast of small clouds can be small and even negative (their ``overdensity'' $(1+\Delta)$ can reach the minimum $(1+\Delta)_u \sim 10^{-2}$ shown in Fig.5 below) they correspond to filaments or underdense objects, surrounded by regions of even lower density, which have decoupled from the expansion of the universe. Hence they do not cool because of adiabatic expansion (or at least this term is much smaller than for the IGM). In particular, we stress that on strongly non-linear scales {\it even objects with a density contrast $\Delta$ equal to $0$ are not described by the linear theory}. Thus, the average density $\rhoa$ loses the significance it has on large scales in the sense that it does not define any longer a density boundary between two physically different regimes. In fact, this role is now played by the density contrast $(1+\Delta)_u$, defined in (\ref{Deltau}), which describes most of the volume of the universe seen at small scales (see Valageas \& Schaeffer 1997). In these regions, patches of matter with the density $\rhoa$ appear as density peaks. This is clearly seen in Valageas et al.(1999b) where a comparison with numerical simulations shows that the {\it same formulation} (\ref{etah}), which is based on a stable-clustering approximation in a statistical sense, provides a reasonable description of objects defined by a density contrast which can vary from $(1+\Delta) = 5000$ downto $(1+\Delta) = 0.5$.

As compared to the original model with only photoionization heating (upper 
figure) the main difference when we include a source term corresponding to 
supernovae (middle figure) or quasars (lower figure) is that the IGM 
temperature keeps increasing until $z \sim 0$ to reach a value $T_{IGM}(z=0) 
\simeq 5 \; 10^5$ K (the decline of $T_{IGM}$ at low $z$ in the original case, 
and for the quasar scenario at $z \sim 0$, is due to adiabatic cooling, 
because of the expansion of the universe). In this case, as seen in (\ref{Tforest}) we take the Lyman-$\alpha$ clouds to be heated to the same temperature in order to have a self-consistent model. However, this high temperature could make it difficult to recover the observed properties of Lyman-$\alpha$ clouds at low $z$. Although this would deserve a detailed study we do not further investigate this point in this article where we mainly consider the energy requirements implied by efficient reheating. Moreover, the opacity due to Lyman-$\alpha$ clouds at low $z$ does not influence much the reionization and reheating history of the universe (the medium is optically thin as shown by the Gunn-Peterson test). Besides, in order to get reliable estimates of the properties of Lyman-$\alpha$ clouds in the case of strong reheating by supernovae or QSOs one would certainly need to take into account the spatial inhomogeneities of this reheating process, which is beyond the scope of the present study. 

In order to obtain the same entropy 
and temperature at $z=0$ for the IGM, we see that the redshift $z_S$ (shown by 
the dashed vertical line), where no just-virialized halo can cool, is higher 
for the quasar scenario (QSO) than for supernova heating (SN). Indeed, we 
obtain:
\beq
\mbox{SN} \; : \; z_S \simeq 0.4 \hspace{1cm} , \hspace{1cm} \mbox{QSO} \; : \; z_S \simeq 2
\label{zSSNQSO}
\eeq
This is due to the peak of the quasar luminosity function at $z \sim 2$ and 
its sharp decline at lower redshift. Indeed, this implies that most of the 
heating process occurs at $z \sim 2$. On the other hand, since the galaxy 
luminosity function evolves more slowly and does not drop at $z \sim 0$, the 
heating process due to stars keeps going on until $z=0$ so that it appears 
delayed as compared to the quasar scenario, see discussion in 
Sect.\ref{Quasars}. This also shows in the behaviour 
of $T_{IGM}$ which keeps strongly increasing until $z=0$ for supernova heating 
while it remains roughly constant (and even shows a slight decline) after 
$z_S$ for quasar heating. The efficiency factors $\alpha_{SN}$ and $\alpha_Q$ 
we use are:
\beq
\left\{ \begin{array}{l} {\displaystyle \mbox{SN} \; : \; \alpha_{SN} = 1 \; , \; \alpha_Q = 0 } \\ \\  {\displaystyle \mbox{QSO} \; : \; \alpha_{SN} = 0  \; , \; \alpha_Q = 0.008 } \end{array} \right.
\label{alphaSNQSO}
\eeq
We can check that they are consistent with the estimates (\ref{TIGMSN}) and 
(\ref{TIGMQ}) and the requirement that $T_{IGM} \sim 5 \; 10^5$ K at $z=0$. 
This latter value for $T_{IGM}$ is due to the constraint $\lag S \rag_{IGM} = 
\log(k T n_b^{-2/3}/ 1 \; \mbox{keV cm}^2) \simeq 2$ which we impose in order 
to get the break at $T \la 1$ keV of the relation $T-L_X$ for clusters, as 
explained above and described in details in Sect.\ref{Clusters}. In fact, for 
supernova heating the efficiency factor $\alpha_{SN} =1$ is not sufficient 
(though not by far since $\alpha_{SN} =1.7$ is enough) to explain the break of 
the cluster $T-L_X$ relation, as discussed in Sect.\ref{Clusters}. However, 
since we must have $\alpha_{SN} \leq 1$ we keep this value for the supernova 
efficiency factor. On the other hand, we note that a very small value for the 
quasar efficiency factor $\alpha_Q$ is sufficient to raise the entropy of the 
IGM to a level high enough to explain the cluster observations. Note also that 
$\alpha_Q \ll \epsilon_Q$, where $\epsilon_Q=0.1$ defined in 
Sect.\ref{Quasars} measures the quasar radiative efficiency. Thus, the quasar 
scenario (QSO) appears to be quite reasonable while the supernova hypothesis 
(SN) seems less likely. However, further work is needed in order to assess with 
a sufficiently good accuracy the efficiency of these two processes of energy 
transfer before one can draw definite conclusions.

\subsection{Entropy production}
\label{Entropy production}

\begin{figure}[htb]

\begin{picture}(230,370)

\epsfxsize=15 cm
\epsfysize=14 cm
\put(-5,-20){\epsfbox{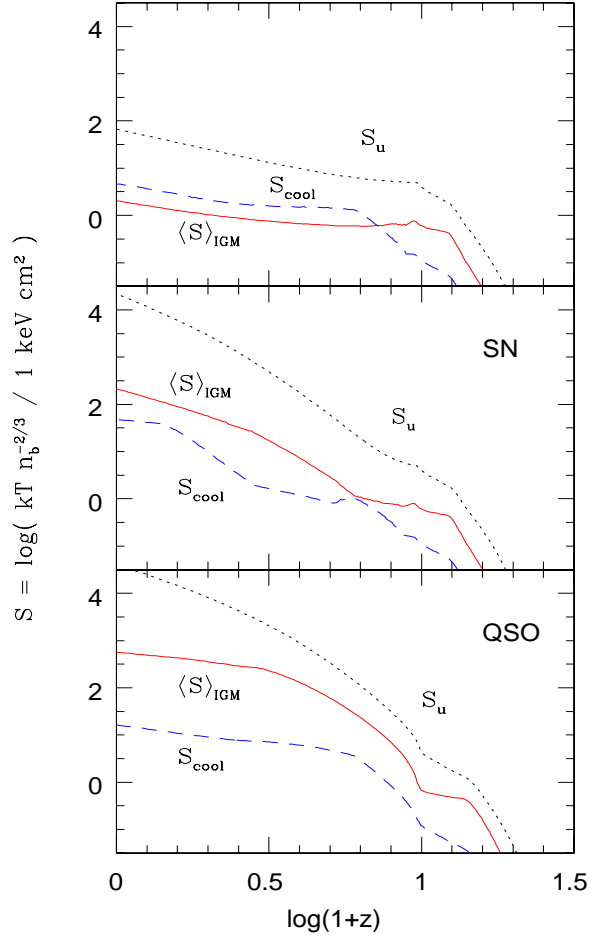}}

\end{picture}

\caption{The redshift evolution of the characteristic entropies of the 
universe. We display the mean IGM entropy $\lag S \rag_{IGM}$ (dot-dashed 
curve), the entropy $S_u$ of underdense regions (solid line) and the entropy 
which would correspond to the smallest galaxy $S_{cool}$ (low dashed line). 
From top downto bottom, the various panels correspond to photoionization 
heating only, supernova heating and quasar heating, as in Fig.\ref{figTO03}.}
\label{figSO03}

\end{figure}

We display in Fig.\ref{figSO03} the redshift evolution of the entropy of the 
gas. It increases with time as structure formation develops and heating 
processes grow. The specific entropy $S_u$ of underdense regions is larger 
than the mean IGM entropy $\lag S \rag_{IGM}$, and increasingly so at smaller 
redshift, because of their low density $n_b$. As was the case for the 
temperature evolution, we note that the entropy shows a faster rise at low $z$ 
for the supernova scenario (SN) than for the quasar heating (QSO). Again this 
is due to the fact that the quasar energy output (proportional to the 
luminosity function) peaks at $z \sim 2$ contrary to the galaxy contribution 
which keeps increasing until $z \la 1$. The entropy $S_{cool}$ is defined 
from the temperature $T_{cool}$ and the density contrast $\Delta_{gal}(z)$. 
The fact that it is lower than $\lag S \rag_{IGM}$, especially at low $z$, 
shows that the smallest galaxies are influenced by the entropy floor set by 
the IGM. However, we recall that $S_{cool}$ is not the entropy of the gas in 
such a galaxy, since the gas compression stops when it reaches the density 
contrast $\Delta_{ad}<\Delta_c$ and it subsequently cools and falls into the 
dark matter potential well, which diminishes its entropy (transfered into the 
radiation).

We note that if we could measure the temperature and the entropy of the gas within Lyman-$\alpha$ clouds or small groups at high redshift $z \ga 1$ we might be able to see which scenario, (SN) or (QSO), is best favoured, using the fact that the redshift evolution is slower for the quasar heating process. However, it is clear that this would not give a definite answer because of the uncertainty associated to these poorly known processes.

\subsection{Time-scales}
\label{Time-scales}

\begin{figure}[htb]

\begin{picture}(230,370)

\epsfxsize=15 cm
\epsfysize=14 cm
\put(-5,-20){\epsfbox{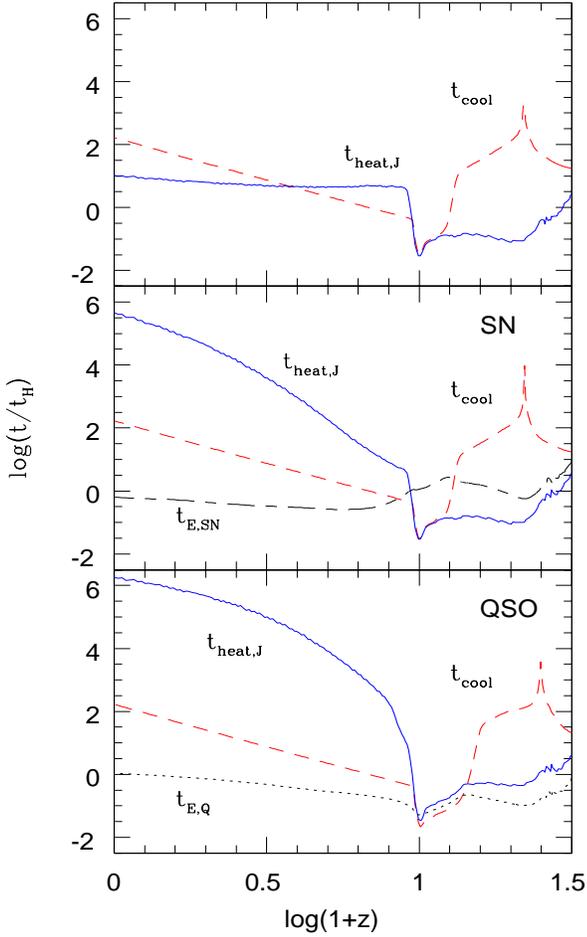}}

\end{picture}

\caption{The redshift evolution of the characteristic heating and cooling time-scales. All times are shown in units of the Hubble time $t_H(z)$. The dashed curve shows the cooling or heating time associated to atomic processes (Compton cooling, collisional excitation,...). The solid line labelled $t_{heat,J}$ shows the photoionization heating time. The dot-dashed (resp. dotted) line labelled $t_{E,SN}$ (resp. $t_{E,Q}$) shows the heating time-scale associated to supernovae (resp. quasars). From top downto bottom, the various panels correspond to photoionization heating only, supernova heating and quasar heating, as in Fig.\ref{figTO03}.}
\label{figtimO03}

\end{figure}

We display in Fig.\ref{figtimO03} the redshift evolution of the various 
characteristic time-scales. For the original model (upper panel), at large 
redshift $8 < z < 24$ the smallest time-scale is the photoionization heating 
time, defined in (\ref{theat}), which means that the IGM temperature increases 
in this redshift range (see Fig.\ref{figTO03}). At lower and larger redshift, 
the smallest time is the Hubble time $t_H(z)$ (corresponding to the ordinate 0 
in the figure) which implies that the IGM cools due to the adiabatic 
expansion. The dashed curve which shows a peak at $\log(1+z) \simeq 1.4$ 
corresponds to atomic processes within the IGM (collisional excitation, 
collisional ionization, bremsstrahlung, Compton cooling or heating,...). Most 
of the time it is dominated by Compton cooling or heating which explains the 
peak at $\log(1+z) \simeq 1.4$ when $t_{Compton}^{-1}=0$ ($T_{IGM}=T_{CMB}$). 
At higher redshift the IGM gas is heated by CMB photons while at lower $z$ the 
IGM is cooled through the interaction with the CMB. Around reionization at 
$z\sim 9$ the main process is collisional excitation (see VS for details). In both lower panels we also display the heating time-scale 
$t_{E,SN}$ or $t_{E,Q}$ associated to the additional energy source. We can see 
that it becomes the dominant process somewhat after reionization at $z \la 8$. 
Then, since $t_E$ becomes the smallest time-scale the corresponding energy 
source heats the IGM up to $T_{IGM} \sim 5 \; 10^5$ K. We can note again that 
the 
supernova heating is somewhat delayed as compared to the quasar scenario. The 
rise at low $z$ of $t_{heat,J}$ as compared to the upper panel is due to the 
growth of the IGM temperature, see (\ref{theat}).

Note that in all three cases the IGM is reheated and reionized by the energy output of galaxies and QSOs (whether it is radiation or kinetic energy). Thus in all scenarios we expect a proximity effect (along a line of sight to a distant QSO the IGM is more ionized close to the quasar) since these processes are not exactly homogeneous (they are more efficient close to the sources). This also holds for the (SN) scenario because QSOs are associated with galactic cores (where a black hole is surrounded by an accretion disk) which means that they are embedded within regions of star formation (which occurs in the host galaxy). In fact, this proximity effect provides a measure of the inhomogeneity of the reheating and reionization process. Note that Davidsen et al.(1996) and Reimers et al.(1997) do not observe any proximity effect which suggests that our homogeneous approximation is reasonable.

\subsection{Reionization history}
\label{Reionization history}

\begin{figure}[htb]

\centerline {\epsfxsize=8.5 cm \epsfysize=6 cm \epsfbox{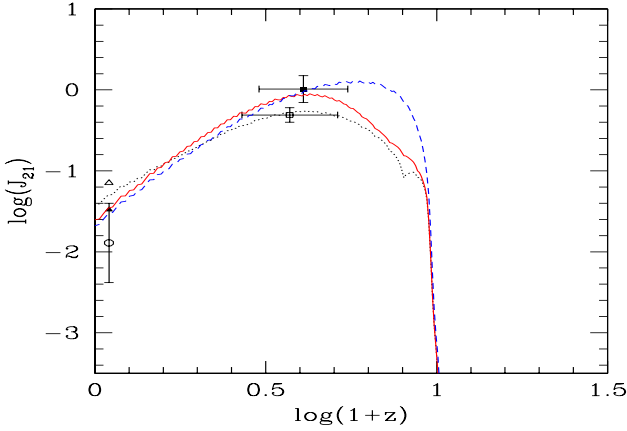}}

\caption{Redshift evolution of the background UV flux $J_{21}$. The dotted line corresponds to photoionization heating only, the solid line to supernova heating and the dashed line to quasar heating.  The data points are from Giallongo et al.(1996) (square), Cooke et al.(1997) (filled square), Vogel et al.(1995) (triangle, upper limit), Donahue et al.(1995) (filled triangle, upper limit) and Kulkarni \& Fall(1993) (circle).}
\label{figJO03}

\end{figure}

We show in Fig.\ref{figJO03} the redshift evolution of the background UV flux $J_{21}$ defined by:
\beq
J_{21} = \frac{ \int J_{\nu} \; \sigma_{HI}(\nu) \; \frac{d\nu}{\nu} }{ \int \sigma_{HI}(\nu) \; \frac{d\nu}{\nu} }  \; / \; 10^{-21} \; \mbox{erg cm}^{-2} \mbox{s}^{-1} \mbox{Hz}^{-1} \mbox{sr}^{-1}
\label{J21}
\eeq
The very sharp rise at $z_{ri} \simeq 9$ corresponds to the reionization 
redshift when the universe suddenly becomes optically thin. We can see that 
all scenarios give nearly the same results since at large redshift the entropy 
of the IGM is not sufficiently large to significantly affect galaxy and quasar 
formation. This was also apparent in Fig.\ref{figTO03} and Fig.\ref{figSO03}. 
In particular, at high $z$ the gas densities are large so that cooling is 
quite easy. As a consequence, the entropy of the IGM cannot prevent the gas 
from cooling and falling into galactic halos to form stars or quasars so that 
{\it the reionization redshift does not depend on the efficiency factors $\alpha$}. 
This is in fact reassuring, since it shows that most of the results we obtained 
in VS (e.g. ionization state of hydrogen and helium,...) 
are still valid and do not depend on the injection of energy into the IGM by 
stars or quasars (the reionization redshift we obtain here is larger 
than in VS because we use $T_{SN} = 10^6$ K instead of
$T_{SN} = 2 \; 10^6$ K). The UV flux in the quasar scenario (QSO) is somewhat larger than for both other cases at $z \sim 7$ ($\log(1+z) \sim 0.9$) although the star formation rate is a bit smaller (see Fig.\ref{figSFRO03} below) because the opacity of the universe is lower. Indeed, this implies a smaller absorption of the radiation emitted by stars and quasars. This is due to the larger temperature of the IGM (see Fig.\ref{figTO03}) which leads to a lower opacity from Lyman-$\alpha$ clouds. However, this is only relevant for $z \ga 5$.

\subsection{Characteristic density contrasts}
\label{Characteristic density contrasts}

\begin{figure}[htb]

\begin{picture}(230,350)

\epsfxsize=15 cm
\epsfysize=14 cm
\put(-5,-35){\epsfbox{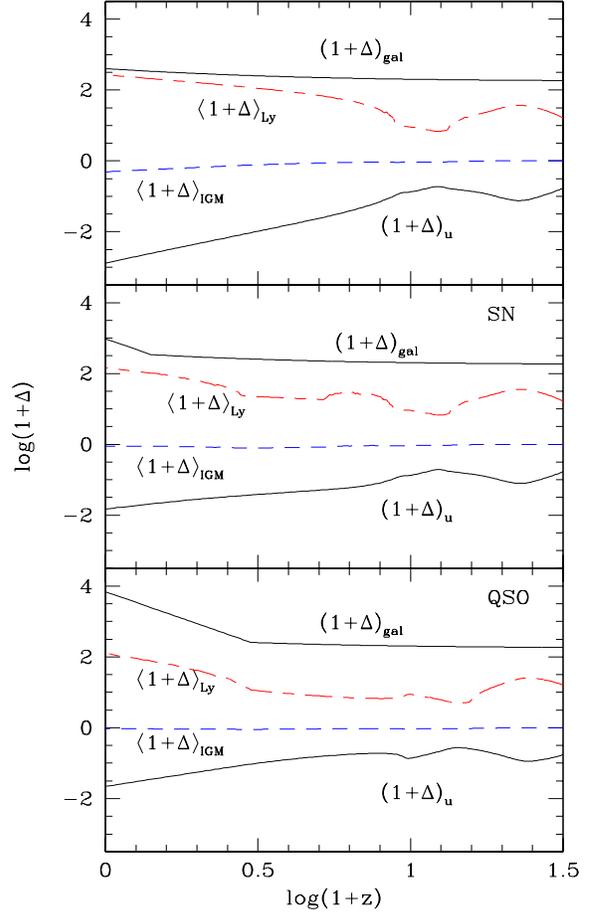}}

\end{picture}

\caption{The redshift evolution of the characteristic density contrasts. We display the density contrast $(1+\Delta)_u$ of large underdense regions in the IGM (lower solid line), $\lag 1+\Delta \rag_{IGM}$ for the mean IGM density (lower dashed line), $\lag 1+\Delta \rag_{Ly}$ for the mass-averaged density within the IGM (upper dot-dashed line) and $(1+\Delta)_{gal}$ for galactic halos (upper solid line).}
\label{figDeltaO03}

\end{figure}

We present in Fig.\ref{figDeltaO03} the redshift evolution of the 
characteristic density contrasts within the universe. The lower solid line 
shows the ``overdensity'' $(1+\Delta)_u$ of the underdense regions which cover 
most of the volume, see (\ref{Deltau}). At large $z$ we have $(1+\Delta)_u 
\simeq 1$ since very few baryonic structures have formed and the universe 
appears as a nearly uniform medium. At low $z$ it declines and can reach very 
low values $(1+\Delta)_u \sim 10^{-2}$ as ``voids'' appear amid filaments and 
halos. It is larger for both lower panels because the IGM temperature is 
higher, which means that the scale $R_d$ is larger, and dark-matter density 
fluctuations are smaller on larger scales (following the behaviour of the 
two-point correlation function). On the other hand, the mean density contrast 
$\lag 1+\Delta \rag_{IGM}$, which takes into account all the matter which is 
not enclosed within galactic halos where the gas was able to cool, from 
``voids'' up to filaments and forest Lyman-$\alpha$ clouds, remains close to 
unity. Indeed, the fraction of matter $1-F_{cool}$ which is not embedded within 
galactic halos remains large until $z=0$ since we have $F_{cool} \la 0.6$ at 
$z=0$, see Fig.\ref{figFracO03}. The mean $\lag 1+\Delta \rag_{Ly}$ is larger than 
$\lag 1+\Delta \rag_{IGM}$ since it is weighted by mass which gives more weight 
to dense regions (filaments, clouds) as compared to low-density areas. 
Finally, the density contrast $\Delta_{gal}$ corresponds to the small galactic 
halos (which are not influenced by the cooling condition $R<R_{cool}$). Thus, 
at all times in the upper panel and at $z \ga 2$ in the lower panels it is 
equal to the ``virialization'' density contrast $\Delta_c(z)$ obtained from 
the usual spherical model. At low redshift in both lower panels it is larger 
than $\Delta_c(z)$ and equal to the expression defined in (\ref{Deltagal}) 
because of entropy considerations. Again, we note that for the quasar scenario 
(QSO) this effect appears somewhat earlier.

\subsection{Compton $y$ parameter}
\label{Compton y parameter}

\begin{figure}[htb]

\begin{picture}(230,380)

\epsfxsize=15 cm
\epsfysize=14 cm
\put(-5,-15){\epsfbox{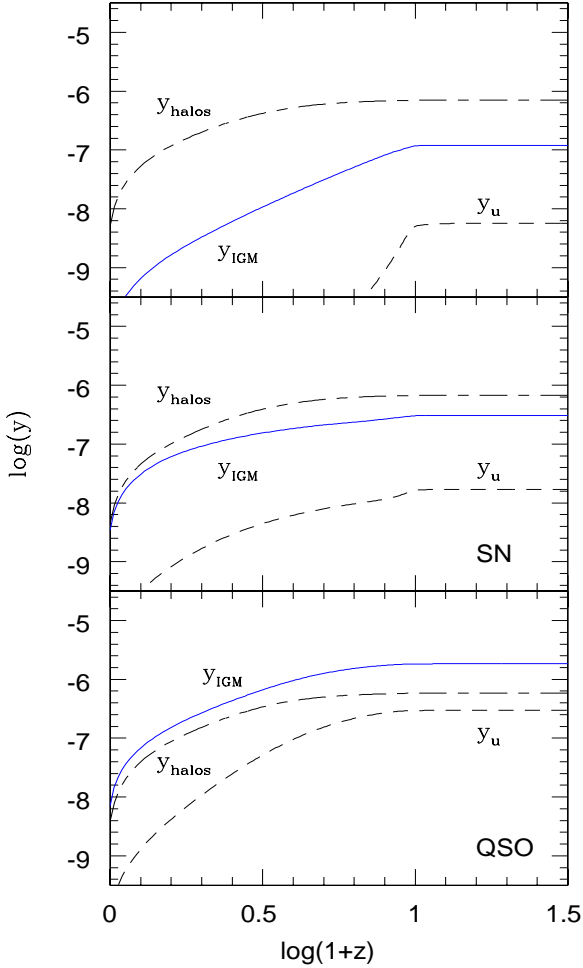}}

\end{picture}

\caption{The Compton parameter $y$ up to redshift $z$ describing the 
Sunyaev-Zeldovich effect from the ``voids'': $y_u$ (lower dashed line), the IGM as a whole (voids, filaments and clouds): $y_{IGM}$ (solid line) and virialized
halos: $y_{halos}$ (upper dashed line).}
\label{figycompO03}

\end{figure}

The hot gas within the IGM or virialized halos (e.g. clusters) scatters some photons of the CMB from the low energy Rayleigh-Jeans part of the spectrum up to the high energy Wien tail. The magnitude of this perturbation is conveniently described by the Compton parameter $y$:
\beq
y(z) = \int_0^z c \frac{dt}{dz} dz \; \sigma_T \; n_e \; \frac{kT}{m_e c^2}
\label{yComp}
\eeq
We consider three components to the global Sunyaev-Zeldovich effect. Thus we define $y_u$ (noted $y_{IGM}$ in VS) describing the effect of low-density regions, $y_{IGM}$ which takes into account all the matter within the IGM (from voids up to filaments and Lyman-$\alpha$ forest clouds) and $y_{halos}$ which describes the effect from virialized halos above $T_{cool}$ (i.e. galaxies and clusters). Both $y_u$ and $y_{IGM}$ reach a plateau at $z_{ri}=9$ since at earlier times the universe was almost exactly neutral. On the other hand $y_{halos}$ saturates earlier because the fraction of matter embedded within massive virialized halos declines at large $z$. The contribution $y_u$ of low-density regions is always small since it only contains a small fraction of matter at low $z$ with a small temperature. Similarly, $y_{halos}$ from galaxies and clusters is much larger than $y_{IGM}$ in the upper panel because although both components contain similar fractions of baryonic matter the temperature of these virialized halos is much larger than the IGM temperature (see Fig.\ref{figTO03}). However, when there is an additional source of energy from supernovae or quasars $y_{IGM}$ becomes non-negligible, especially for the quasar scenario (QSO) where $T_{IGM}$ rises earlier. Note that on general grounds we can expect $y_{IGM}$ to be at most of the same order as $y_{halos}$ since a large fraction of matter is embedded within virialized halos and the temperature of the IGM should not be larger than $10^6$ K while the temperature of these collapsed objects can be much larger (e.g. clusters have $T \sim 10^7$ K). A more precise study of the Compton parameter induced by clusters is presented in Valageas \& Schaeffer (1999b). In any case, we find that the Compton parameter from any contribution is much lower than the COBE/FIRAS upper limit $y < 1.5 \; 10^{-5}$ provided by observations (Fixsen et al.1996). This means that we cannot constrain the entropy scenario from its effect on the Compton parameter using current observations. On the other hand, it shows that we do not contradict the data, hence these heating processes remain plausible explanations for the cluster observations. 

We can also compute the X-ray background provided by the IGM, from underdense regions up to Lyman-$\alpha$ clouds in filaments. However, we find that it is negligible since we get at most (for the QSO scenario) a flux in the 0.5-2 keV band of $\Phi \sim 10^{-17}$ erg s$^{-1}$ cm$^{-2}$ deg$^{-2}$ from the IGM, and $\Phi \sim 10^{-13}$ erg s$^{-1}$ cm$^{-2}$ deg$^{-2}$ from virialized halos which have not cooled, as compared to the observed extragalactic intensity $I \sim 8 \; 10^{-12}$ erg s$^{-1}$ cm$^{-2}$ deg$^{-2}$ (Miyaji et al.1998). On the other hand, the contribution from galaxies and clusters is of the order of $10^{-11}$ erg s$^{-1}$ cm$^{-2}$ deg$^{-2}$. Indeed, the X-ray flux is very sensitive to the density (it varies as $n_e^2$) and to the temperature. Thus, most of the contribution comes from high temperature clusters $T \ga 0.5$ keV while the low temperature of the IGM $T_{IGM} \sim 5 \; 10^5$ K $\sim 0.08$ keV implies a very small contribution since $\exp(-0.5/0.08) \sim 10^{-3}$.

\section{Feedback on structure formation}
\label{Feedback on structure formation}

\subsection{Star formation}
\label{Star formation}

As described in Sect.\ref{Effect of the IGM entropy on galaxy formation} the entropy of the IGM inhibits the cooling of the gas which in turns decreases the efficiency of star formation. In particular, after the redshift $z_S$ the gas which would become embedded within new non-linear structures cannot cool within a few Hubble times at formation because its entropy prevents its density to reach large enough values to start cooling efficiently. Thus, in our model we obtain a population of ``old'' galaxies which gradually convert their matter content into stars but the overall fraction of gas which can cool does not increase any more. After a while, when a large part of this matter has formed stars, this leads to a decrease of the star formation rate at low $z$. Note that we neglect cooling flows within groups and clusters which provide an additional source of cool gas which may form stars. However, this feedback effect onto galaxy formation is likely to persist in a more detailed model.

\begin{figure}[htb]

\centerline {\epsfxsize=8.5 cm \epsfysize=6 cm \epsfbox{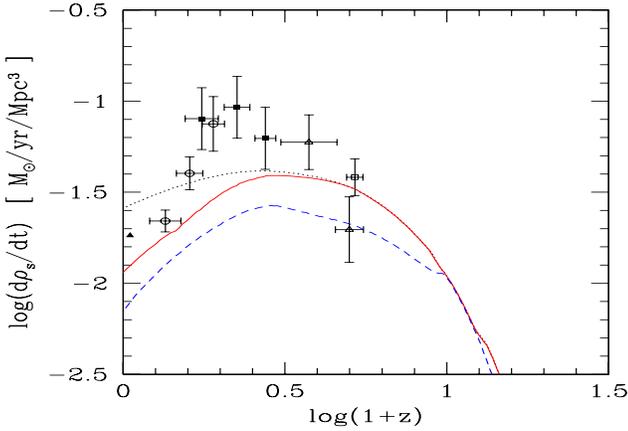}}

\caption{The redshift evolution of the comoving star formation rate. The dotted line corresponds to photoionization heating only, the solid line to supernova heating (SN) and the dashed line to quasar heating (QSO). The data points are from Madau (1999), see references therein.}
\label{figSFRO03}

\end{figure}

\begin{figure}[htb]

\begin{picture}(230,370)

\epsfxsize=15 cm
\epsfysize=14 cm
\put(-5,-25){\epsfbox{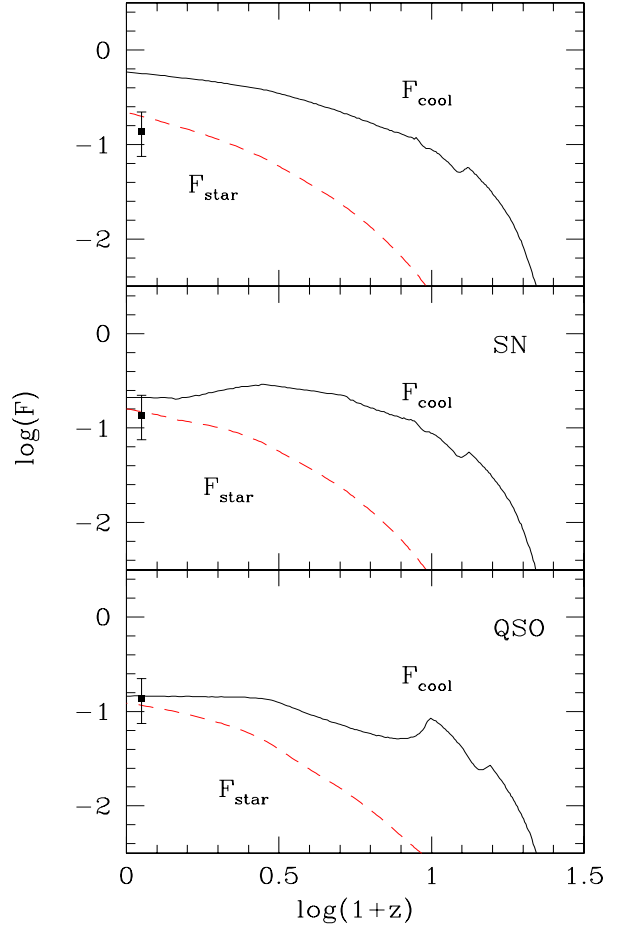}}

\end{picture}

\caption{The redshift evolution of the fraction of baryonic matter embedded within cooled objects $F_{cool}$ (solid line) and stars $F_{star}$ (dashed line). The data point at $z \simeq 0$ shows the observed mass within stars, from Fukugita et al.(1998).}
\label{figFracO03}

\end{figure}

We show in Fig.\ref{figSFRO03} the comoving star formation rate we obtain for 
the three scenarios. We clearly see the inhibition of star formation as 
compared to the case with photoionization heating only. This also shows at 
large redshift $z>z_S$, where new galaxies appear but the high entropy of the 
gas prevents the formation of the smallest galaxies which occur in the case 
with $\alpha_{SN}=\alpha_Q=0$. As explained in Sect.\ref{Effect of the IGM entropy on galaxy formation} this feedback effect is larger at low $z$ where structure formation is further developped (hence the energy source is high) and the baryonic density is lower (hence cooling is less efficient). Moreover, this feedback effect onto star 
formation may partly explain the sharp drop of the star formation rate 
observed at low redshift $z<1$. Indeed, this decline is not very easy to 
obtain in usual models since at $z \sim 0$ the fraction of baryons which has 
been converted into stars is still very small $F_{star} \sim 10 \%$. As a 
consequence, it is difficult to get a sudden stop of the star formation 
process because there is plenty of gas available (although one can obtain in a 
natural fashion a decline as shown by the dotted line). On the other hand, the 
supernova or quasar heating of the IGM is able to suddenly change the 
conditions of star formation when the entropy of the gas becomes of the order 
of the entropy generated by gravitational collapse within the new non-linear 
halos. 

Although the shape of the redshift evolution of the comoving star 
formation rate we obtain for the quasar scenario (dashed line) agrees with 
observations, its normalization is somewhat too low. This might be the sign of 
a shortcoming of our description, in particular the assumption that reheating 
is uniform could lead to an overestimate of the entropy feedback onto star 
formation, since in a more realistic model with inhomogeneous reheating this 
effect may require more time to affect all the gas (especially for the small 
galaxies which form far away from clusters or groups of large galaxies and 
quasars). In fact, the star formation rate for the photoionization only case is already a bit two low at $z \sim 1$. Thus our model of star formation is not perfect yet and we might underestimate the supernovae heating by a factor 2. Although this would translate into a supernovae efficiency factor smaller than unity, $\alpha_{SN} \simeq 0.8$, this value remains quite large and it does not modify our conclusions (e.g. that the IGM is more likely to have been reheated by quasars). Thus, we think these results are already quite encouraging, in 
view of the simplicity of our model. On the other hand, note that we obtained 
a correct amplitude for the UV flux as shown in Fig.\ref{figJO03}.

We show in Fig.\ref{figFracO03} the fraction of matter enclosed within cooled 
objects (galaxies) and stars ($F_{star} = \Omega_{star}/\Omega_b$). We see 
that our results agree with the observed mass of stars (which may be more 
robust than the observed redshift evolution of the star formation rate itself). 
This could be expected since our model for galaxy formation (though in a more 
detailed version) was already checked in Valageas \& Schaeffer (1999a) against 
observations. In particular, the cooled fraction $F_{cool}$ at $z=0$ obtained with photoionization heating only (upper panel) is rather large. However, the mass of stars we obtain agrees with observations which shows that we do not encounter the overcooling problem. This is due to the small star formation efficiency implied by several processes: ejection of matter by supernovae and stellar winds, energy released by halos mergings and collapse.

Of course, we recover the behaviour seen in Fig.\ref{figSFRO03}. 
In particular, {\it for strong heating of the IGM we clearly see the saturation of 
the mass of cooled gas at low $z$ and the upper bound for the mass of stars it 
implies}. Note that this feedback effect is quite general. Indeed, we require 
the reheating of the IGM to be large enough to affect cluster formation at 
$z \sim 0$ for halos with a virial temperature $T \sim 0.5$ keV. It is clear 
that this implies a significant effect onto galaxy formation at low redshift, 
since galaxies consist of shallower potential wells $T \la 0.1$ keV. Since we 
must {\it simultaneously} describe galaxies, quasars and clusters, this leads 
to contradictory constraints. Indeed, the cluster $T-L_X$ relation requires a 
high reheating temperature $T \ga 5 \; 10^5$ K (see Sect.\ref{Clusters}) while 
the observed star formation rate requires a small enough reheating, $T \la 5 
\; 10^5$ K, so as not to inhibit too much galaxy formation. Thus, we see from 
Fig.\ref{figSFRO03} and Fig.\ref{figclusTLO03} that these constraints imply a 
reheating temperature $T \sim 5 \; 10^5$ K and $\alpha_Q \sim 0.008$. This 
clearly shows the importance of using global models like ours (even though 
simplified) which allow one to evaluate the consequences of such processes on 
all objects (galaxies, clusters,...). Indeed, this provides strong constraints 
on such descriptions and it is the only way to test the {\it global} validity 
of these scenarios which should {\it simultaneously} account for all structure 
formation processes. We also note that a large fraction of baryonic matter at
$z \sim 0$ is embedded within density fluctuations in the IGM and 
Lyman-$\alpha$ clouds with a rather large temperature $T \ga 5 \; 10^5$ K.
Part of this component may be difficult to observe.

\subsection{Quasar luminosity function}
\label{Quasar luminosity function}

\begin{figure}[htb]

\begin{picture}(230,470)

\epsfxsize=26 cm
\epsfysize=18 cm
\put(-28,-30){\epsfbox{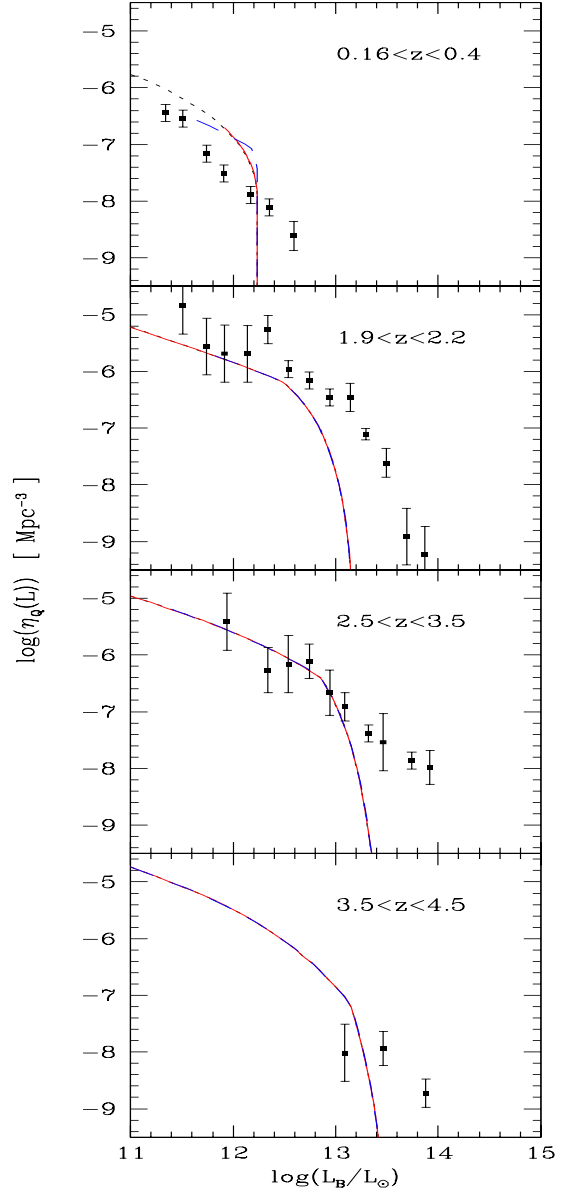}}

\end{picture}

\caption{The evolution with redshift of the B-band quasar luminosity function in comoving Mpc$^{-3}$. The dotted lines correspond to photoionization heating only, the solid lines to supernova heating and the dashed lines to quasar heating. The data points are from Pei (1995).}
\label{figquasO03}

\end{figure}

We display in Fig.\ref{figquasO03} the redshift evolution of the quasar 
luminosity function for all scenarios. We can see that at large redshifts 
$z \ga 2$ all curves nearly superpose since the additional energy source 
(from quasars or supernovae) plays no role at early times. On the other hand, 
as was the case for star formation, at small redshift $z \la 0.5$ the 
entropy ``floor'' induced by quasar heating leads to a decrease of the quasar 
luminosity function which slightly improves the agreement with observations. 
Thus, although the decline at low $z$ of the QSO multiplicity function is a 
natural outcome of our model even for the original scenario with 
photoionization heating only (dotted line), part of this decrease may also be 
due to the entropy production by quasars at earlier times.  In view of the 
simplicity of our model for quasar formation, which is a natural outcome of our
description of galaxies, we think our predictions agree reasonably well with 
observations. Of course, the physics of quasars may be more intricate than
the description we use in this article but any meaningfull improvement would 
require a detailed model of the accretion processes leading to quasar formation,
which is beyond the scope of this study.

\subsection{Clusters}
\label{Clusters}

As we noticed in the introduction, the reheating of the IGM by supernovae or 
quasars at $z \ga 1$ can affect the formation of clusters at lower redshifts 
$z \la 0.5$ since it leads to a mimimum entropy $\lag S \rag_{IGM}$ of the gas 
which can break the expected scaling of the relation temperature - X-ray 
luminosity of 
clusters. In order to obtain an estimate of the characteristic virial 
temperature $T_{ad,cl}$ where this transition should occur we define:
\beq
T_{ad,cl} = T_{IGM} \; \left( \frac{1+\Delta_c}{1+\Delta_u} \right)^{2/3} \; 10^{\lag S \rag_{IGM} - S_u}
\label{Tadcl}
\eeq
where we used (\ref{Deltaad}). This is the temperature of the gas at the density contrast $\Delta_c(z)$ 
(which defines clusters in our model) with the mean entropy of the IGM 
$\lag S \rag_{IGM}$. Thus, massive clusters with $T \gg T_{ad,cl}$ are not 
affected by the entropy floor of the IGM because shock heating during the 
gravitational collapse of the halo generates a much larger entropy so that the 
usual scaling law is recovered. On the other hand, within smaller potential 
wells the gas has a larger entropy than the one produced by shock heating 
which leads to a smoother gas density profile and to a smaller density. This 
implies a smaller luminosity since $L_X \propto n_e^2$. In (\ref{Tadcl}) we used the virial density rather than the core density, which is significantly larger, because most of the mass is characterized by densities of order $\Delta_c$ and we assume isothermal equilibrium. However, it is clear that (\ref{Tadcl}) is only a rough estimate which could be uncertain by a factor 2.

\begin{figure}[htb]

\centerline {\epsfxsize=8.5 cm \epsfysize=6 cm \epsfbox{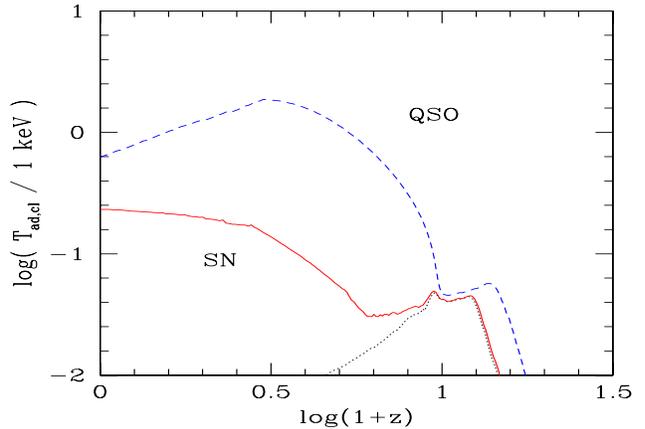}}

\caption{Redshift evolution of the characteristic temperature $T_{ad,cl}$ which describes the influence of reheating on cluster formation. The dotted line corresponds to photoionization heating only, the solid line to supernova heating and the dashed line to quasar heating.}
\label{figTclusO03}

\end{figure}

We show in Fig.\ref{figTclusO03} the redshift evolution of $T_{ad,cl}$. Again, 
we can check that in the quasar scenario (QSO) the heating of the IGM occurs 
earlier than for the supernova case (SN). Moreover, since the quasar 
luminosity function drops at low $z$ so that adiabatic cooling becomes the 
main process, as seen in Fig.\ref{figtimO03}, $T_{ad,cl}$ decreases at low 
$z$. On the other hand, for the supernova scenario the fact that the redshift 
$z_S$ is smaller leads to a smoother evolution at $z \sim 0$. In any case, we 
see that we obtain at small redshift a characteristic temperature $T_{ad,cl} 
\sim 0.5$ keV. This is similar to the values which are used in studies of 
cluster formation (Cavaliere et al.1997 use $T_{ad,cl} \simeq 0.5 - 0.8$ keV 
while Valageas \& Schaeffer 1999b use $T_{ad,cl} \simeq 0.35$ keV). Indeed, 
this is how we chose the parameters $\alpha$. Of course, smaller efficiency 
factors than those set in (\ref{alphaSNQSO}) lead to smaller IGM entropy and 
lower $T_{ad,cl}$. However, it is interesting to note that a larger $\alpha_Q$ 
does not change much our results at $z=0$ for the quasar scenario (QSO) because 
of the feedback effect we described above. On the other hand, for the supernova 
scenario (SN) the available range is already limited by the upper bound 
$\alpha_{SN} \le 1$.

From the mean entropy of the IGM, or the characteristic temperature $T_{ad,cl}$, we can obtain the cluster temperature - X-ray luminosity relation, as in Valageas \& Schaeffer (1999b). The bolometric X-ray luminosity $L_X$ of a cluster of volume $V$ is:
\beq
L_{bol} = \int_V n_e^2 \Lambda_b(T_g) dV
\label{Lbol1}
\eeq
where $n_e$ is the electron number density and $\Lambda_b(T_g)$ is the bremsstrahlung emissivity function (in erg cm$^3$ s$^{-1}$) at temperature $T_g$. Thus, contrary to the Sunyaev-Zel'dovich effect, see (\ref{yComp}), the X-ray luminosity strongly depends on the density profile of the hot gas within the cluster. During the gravitational collapse of the cluster shocks heat the gas up to the virial temperature $T$ of the dark matter halo. However, the adiabatic compression of the gas from the IGM also heats the gas up to $T_{ad,cl}$ which is a lower bound for the gas temperature $T_g$. In order to take into account both of these effects, we simply write for the final temperature of the gas:
\beq
T_g = T + T_{ad,cl}
\label{Tgas}
\eeq
Next, in order to obtain the density profile of the gas within the dark matter potential well we assume isothermal and hydrostatic equilibrium at the temperature $T_g$ and we obtain (see also Cavaliere \& Fusco-Femiano 1978):
\beq
\rho_g \propto \rho^{-\beta} \propto r^{-2\beta} \hspace{0.5cm} \mbox{with} \hspace{0.5cm} \beta = \frac{T}{T_g} = \frac{T}{T + T_{ad,cl}}
\label{beta}
\eeq
where we used an isothermal profile $\rho \propto r^{-2}$ for the dark matter halo. Thus, for deep potential wells $T \gg T_{ad,cl}$ we have $\beta \simeq 1$ and the gas follows the dark matter density profile while for cool clusters $T \la T_{ad}$ we get $\beta \la 1$ and the gas density profile is smoother than the dark matter distribution. This change of the shape of the gas density profile leads to a break in the relation $T_g-L_X$. Moreover, in the inner parts of the cluster the density is large enough to lead to a small cooling time so that a cooling flow develops and some of the gas forms a cold component which does not emit in X-ray any longer. Thus, we define the cooling radius $R_c$ as the point where the gas density reaches the threshold $\rho_{gc}$ such that:
\beq
t_{cool} = t_H \hspace{0.5cm} \mbox{with} \hspace{0.5cm} t_{cool} = \frac{3 \mu_e^2 m_p k T_g}{2\mu \rho_{gc} \Lambda_c(T_g)}
\label{tcoolclus}
\eeq
where $\Lambda_c(T_g)$ is the cooling function (which is dominated by bremsstrahlung cooling for $T>1$ keV) and $t_H(z)$ is the Hubble time. At large radii $r>R_c$ the density is lower than $\rho_{gc}$ hence the local cooling time is larger than the Hubble time. Then, the gas distribution and the temperature had not had time to evolve much and the X-ray emissivity is proportional to $\rho_g^2 \; \Lambda_b(T_g)$, see (\ref{Lbol1}). On the other hand, within the cooling radius $R_c$ the gas had time to cool and form dense cold clouds. However, we consider that some of the gas is still hot and emits in X-ray as cooling does not proceed in a uniform fashion (Nulsen 1986; Teyssier et al.1997). The density of this remaining gas has to be of order $\rho_{gc}$ and we write the X-ray luminosity of the cluster as:
\beq
\begin{array}{l} {\displaystyle L_X = \epsilon \; \left( \frac{\rho_{gc}}
{\mu_e m_p} \right)^2 \; \Lambda_b(T_g) \; \frac{4 \pi R_c^3}{3} } \\ \\ 
{\displaystyle \hspace{2cm} \times \left\{ 1 + \frac{3}{4\beta-3} \left[ 
1 - \left( \frac{R_c}{R_g} \right)^{4\beta-3} \right] \right\} } \end{array}
\eeq
where the factor $1$ describes the contribution of the core, within $R_c$, and 
the second term comes from the halo (note that the contribution within $R_c$ is
never much larger than the one from $r \ga R_c$). The factor $\epsilon=3$ is 
a parameter of 
order unity which we use to normalize our relation to observations for massive 
clusters ($T_g > 1$ keV). We expect $\epsilon \ga 1$ which is indeed the case. 
It measures the density fluctuations of the gas distribution, since $L_X 
\propto \lag n_e^2 \rag$ and at any radius $\lag n_e^2 \rag \ga \lag n_e 
\rag^2$. Note that our description is similar to the model developed by 
Cavaliere et al.(1997,1998) to describe the relation between the gas and the 
dark matter density profile, which shows in the quantity $\beta$. However, 
while they define a core radius from the density distribution itself (because 
they use a dark matter profile which grows more slowly than $r^{-2}$ at small 
$r$) the core radius we use describes the cooling of the gas, independently of 
the profile of the underlying dark matter halo. Using a shallower dark matter
density profile would give similar results with a slightly larger $T_{ad,cl}$
and $\alpha_Q$ (see also Valageas \& Schaeffer 1999b).

\begin{figure}[htb]

\centerline {\epsfxsize=8 cm \epsfysize=11 cm \epsfbox{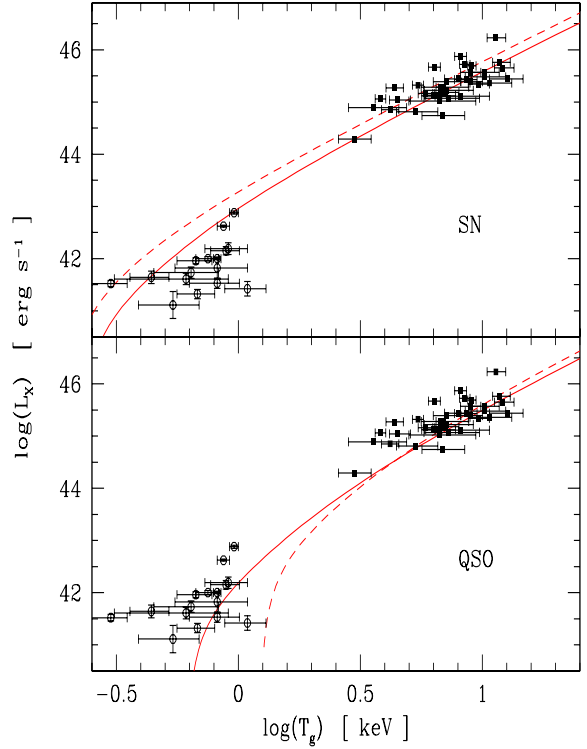}}

\caption{The cluster temperature - X-ray luminosity relation at $z=0$ (solid lines) and $z=1$ (dashed lines). The upper panel shows the supernova scenario (SN) and the lower panel the quasar scenario (QSO). The data points are from Mushotzky \& Scharf (1997) for clusters and from Ponman et al.(1996) for groups.}
\label{figclusTLO03}

\end{figure}

We show our results in Fig.\ref{figclusTLO03} at $z=0$ and $z=1$, for both 
(SN) and (QSO) scenarios, using the redshift evolution we obtained in 
Fig.\ref{figTclusO03} for $T_{ad,cl}$. First, we can check that our results 
agree with observations for hot clusters $T > 1$ keV. Then, we see that the 
initial entropy of the IGM leads to a break of the relation $T_g - L_X$ at low 
temperatures. However, for the supernova heating scenario we would require 
$\alpha_{SN} \simeq 1.7$ to get a sufficiently large effect (though even for 
$\alpha_{SN} =1$ we already see a knee in the relation $T-L_X$). On the other 
hand, for the quasar heating case we need $\alpha_Q \simeq 0.008$. Note that 
the strong redshift evolution of $T_{ad,cl}$ in the quasar scenario, seen in 
Fig.\ref{figTclusO03}, leads to a clear redshift dependence of the break of 
the relation $T_g - L_X$. On the contrary, the slow redshift evolution of 
$T_{ad,cl}$ in the supernova scenario leads to a smoother redshift dependence 
of the relation $T_g - L_X$. This suggests that observations of the evolution 
of the temperature - luminosity relation at low $L_X$ could provide some 
contraints on the (QSO) scenario. On the other hand, at large $L_X$ there is 
almost no redshift evolution, which is consistent with observations 
(Mushotzky \& Scharf 1997). Finally, we note that we may underestimate the 
effect of the heating from supernovae or quasars onto cluster formation in our 
model. Indeed, while we assumed this energy source to be homogeneous the gas 
which will later build a cluster is more likely to be reheated than an average 
calculation would show since clusters form at density peaks where the local 
density of galaxies and quasars is higher than average. This means that the 
break of the relation $T_g - L_X$ could appear at a slightly larger temperature 
than shown in Fig.\ref{figTclusO03}. This might ``help'' the supernova scenario 
as an efficiency factor $\alpha_{SN}$ smaller than unity could be sufficient. 
However, it would probably remain close to $\alpha_{SN}=1$. On the other hand, 
we note that our constraints for the reheating of the IGM do not depend much on 
the details of the model of clusters since in any case in order to get a break 
of the $T_g - L_X$ relation at $T_g \la 1$ keV one necessarily needs to 
introduce a characteristic temperature $T_{ad,cl} \sim 0.5$ keV 
(Fig.\ref{figTclusO03}) which sets the location of this bend. A more detailed 
model of clusters is presented in Valageas \& Schaeffer (1999b) where a good 
match with observations is obtained with $T_{ad,c} \simeq 0.35$ keV.

\section{Conclusion}

In this article we have described an analytic model for structure formation 
processes which deals in a consistent fashion with quasars, galaxies, 
Lyman-$\alpha$ absorbers and underdense regions within the IGM. This allows us 
to obtain the reheating and reionization history of the universe, as well as 
the evolution of the entropy of the gas. We considered three scenarios, with 
different efficiency factors for the transfer of energy from supernovae or 
quasars into the IGM. Thus, we have shown that the energy provided by quasars 
is sufficient to reheat the universe and raise the mean entropy of the IGM up 
to the value required to match the ``floor'' level observed in cool clusters. 
This is an upper bound on the entropy production and this value allows to 
explain the behaviour of the cluster $T-L_X$ relation. On the other hand, the 
supernova heating scenario would require an efficiency factor of order
unity ($\alpha_{SN} \simeq 1.7$). Thus, {\it the IGM is more likely to have 
been reheated by quasars than by supernovae}. 

Of course, a more realistic treatment would account for the details of the 
quasar interaction with their gaseous environment. However, our study already
shows that quasar-driven outflows can provide an important heating mechanism.
On the other hand, a detailed model of the inhomogeneous character of this
reheating is probably necessary in order to evaluate its effects on 
Lyman-$\alpha$ clouds since most of the opacity may come from clouds located
far away from quasars which have not been reheated. 

We showed that {\it the feedback of entropy production onto structure formation may 
partly account for the decline at low $z$ of the comoving star formation rate 
and of the quasar luminosity function, in addition to cluster observations}. 
This is an interesting prospect since it links different processes to the same 
phenomenon and it gives additional weight to the hypothesis of such a 
reheating scenario. Moreover, it provides a narrow range for the reheating of 
the IGM ($T \sim 5 \; 10^5$ K) since we must satisfy the contradictory 
constraints arising from clusters (which require a large reheating so as to modify the $T-L_X$ relation) and from galaxies and quasars (which require a small reheating so that galaxy formation is not too much inhibited). Thus, it is important to {\it simultaneously} address these processes in order to check the validity of a given scenario. On the other hand, the reionization process of the universe is almost not modified which means that our results for the ionization state of the gas and the background UV flux should be quite robust.

Then, we found that although both scenarios, normalized to the current value 
of the entropy measured in cool clusters, are very similar, the reheating due 
to quasars occurs a bit earlier ($z_S \sim 2$) than for supernovae ($z_S 
\simeq 0.4$) because of the sharp drop at $z<2$ of the quasar luminosity 
function. This might give a clue to discriminate these two sources of energy 
from observations. However, it is clear that further work is needed in order 
to get stronger constraints on the possible efficiency of these reheating 
processes, for instance through very detailed numerical simulations. 
Nevertheless, despite the small discrepancies we get for the star formation 
rate or the quasar luminosity function as compared with observations, it 
appears quite remarkable that a simple analytic model like ours, which 
necessarily involves some approximations (e.g. we do not take into account the 
scatter of the galaxy or quasar properties, nor the inhomogeneity of the 
supernova or quasar heating), provides such a good description of structure 
formation processes. Indeed, we note that at $z \sim 0$ we describe objects 
which span a wide range in density, from $(1+\Delta) \sim 10^{-2}$ for voids 
and low-column density Lyman-$\alpha$ absorbers up to $(1+\Delta) \sim 10^3$ 
for old galaxies, as well as in mass, from $10^9 M_{\odot}$ for Lyman-$\alpha$ 
clouds up to $10^{15} M_{\odot}$ for clusters. The fact that we can build a 
unified consistent model for this broad variety of structures strongly suggests 
that hierarchical scenarios like ours, with adequate models for galaxy 
formation and radiative processes, provide a realistic description of the 
actual universe.

\end{document}